\documentclass[fleqn,usenatbib]{mnras}

\usepackage{newtxtext,newtxmath}

\usepackage[T1]{fontenc}

\DeclareRobustCommand{\VAN}[3]{#2}
\let\VANthebibliography\thebibliography
\def\thebibliography{\DeclareRobustCommand{\VAN}[3]{##3}\VANthebibliography}

\usepackage{subcaption}
\usepackage{amsmath}

\usepackage{graphicx}	
\usepackage{amsmath}	
\usepackage{bm}         
\usepackage{multirow}   
\usepackage{tikz,xcolor,hyperref} 
\usepackage{enumitem}  
\usepackage{caption}

\newcommand{\xmm}{{\em XMM}} 
\newcommand{\chandra}{{\em Chandra}}
\newcommand{\erosita}{{\em eROSITA}}
\newcommand{\xga}{{\sc Xga}}

\newcommand{\daxa}{{\sc Daxa}}
\newcommand{\esass}{{\tt eSASS}}
\newcommand{\xapa}{{\sc Xapa}}
\newcommand{\democlusterfull}{Abell 3158}
\newcommand{\democluster}{A3158}
\newcommand{\erass}{{eRASS1}}
\newcommand{\rad}{{$R_{500}$}}

\newcommand{\migkasNORM}{0.99$\pm$0.04}
\newcommand{\migkasSCATTER}{0.11$\pm$0.05}
\newcommand{\migkasSAMP}{68}
\newcommand{\migkasSLOPE}{1.05$\pm$0.07}

\newcommand{\bulTNORM}{{1.00$\pm$0.07}}
\newcommand{\bulTSCATTER}{{0.07$\pm$0.05}}
\newcommand{\bulTSAMP}{{30}}
\newcommand{\bulTSLOPE}{{0.9$\pm$0.1}}

\newcommand{\bulLNORM}{{0.95$\pm$0.03}}
\newcommand{\bulLSCATTER}{{0.19$\pm$0.03}}
\newcommand{\bulLSAMP}{{76}}
\newcommand{\bulLSLOPE}{{1.06$\pm$0.03}}

\newcommand{\LsoftSAMP}{{73}}
\newcommand{\SAMPT}{{37}}
\newcommand{\LbolSAMP}{{74}}

\newcommand{\LsoftNORMfix}{{1.19$\pm$0.06}}
\newcommand{\LsoftSCATTERfix}{{0.39$\pm$0.04}}
\newcommand{\LsoftNORMfixT}{{1.21$\pm$0.08}}
\newcommand{\LsoftSCATTERfixT}{{ 0.37$\pm$0.05}}
\newcommand{\LsoftNORM}{{0.95$\pm$0.04}}
\newcommand{\LsoftSLOPE}{{0.86$\pm$0.04}}
\newcommand{\LsoftSCATTER}{{0.36$\pm$0.04}}
\newcommand{\LsoftNORMT}{{0.97$\pm$0.05}}
\newcommand{\LsoftSLOPET}{{0.82$\pm$0.05}}
\newcommand{\LsoftSCATTERT}{{0.32$\pm$0.04}}

\newcommand{\LbolNORMfix}{{1.34$\pm$0.07}}
\newcommand{\LbolSCATTERfix}{{0.43$\pm$0.05}}
\newcommand{\LbolNORM}{{0.77$\pm$0.04}}
\newcommand{\LbolSLOPE}{{0.87$\pm$0.05}}
\newcommand{\LbolSCATTER}{{0.41$\pm$0.04}}
\newcommand{\LbolNORMfixT}{{1.32$\pm$0.09}}
\newcommand{\LbolSCATTERfixT}{{0.38$\pm$0.05}}
\newcommand{\LbolNORMT}{{0.78$\pm$0.05}}
\newcommand{\LbolSLOPET}{{0.82$\pm$0.05}}
\newcommand{\LbolSCATTERT}{{0.34$\pm$0.05}}

\newcommand{\TNORMfix}{{1.28$\pm$0.07}}
\newcommand{\TSCATTERfix}{{0.19$\pm$0.05}}

\newcommand{\TNORM}{{0.98$\pm$0.06}}
\newcommand{\TSLOPE}{{1.0$\pm$0.2}}
\newcommand{\TSCATTER}{{0.20$\pm$0.05}}

\newcommand{\RICHLSAMP}{{73}}
\newcommand{\RICHTSAMP}{{37}}
\newcommand{\eRICHLsoftNORM}{{0.74$\pm$0.07}}
\newcommand{\eRICHLsoftSLOPE}{{1.3$\pm$0.2}}
\newcommand{\eRICHLsoftSCATTER}{{0.77$\pm$0.07}}
\newcommand{\eRICHLsoftNORMT}{{0.8$\pm$0.1}}
\newcommand{\eRICHLsoftSLOPET}{{1.3$\pm$0.3}}
\newcommand{\eRICHLsoftSCATTERT}{{0.8$\pm$0.1}}
\newcommand{\xRICHLsoftNORM}{{0.87$\pm$0.07}}
\newcommand{\xRICHLsoftSLOPE}{{1.4$\pm$0.2}}
\newcommand{\xRICHLsoftSCATTER}{{0.69$\pm$0.06}}
\newcommand{\xRICHLsoftNORMT}{{1.0$\pm$0.1}}
\newcommand{\xRICHLsoftSLOPET}{{1.2$\pm$0.2}}
\newcommand{\xRICHLsoftSCATTERT}{{0.70$\pm$0.09}}

\newcommand{\eRICHTNORM}{{0.96$\pm$0.06}}
\newcommand{\eRICHTSLOPE}{{0.5$\pm$0.1}}
\newcommand{\eRICHTSCATTER}{{0.20$\pm$0.07}}

\newcommand{\xRICHTNORM}{{1.3$\pm$0.06}}
\newcommand{\xRICHTSLOPE}{{0.53$\pm$0.09}}
\newcommand{\xRICHTSCATTER}{{0.26$\pm$0.03}}

\defcitealias{Whelan2022}{W22}
\defcitealias{Migkas2024}{M24}
\defcitealias{Bulbul2024}{B24}
\defcitealias{turner2025}{T25}
\defcitealias{Turner2022}{T22}

\definecolor{lime}{HTML}{A6CE39}
\DeclareRobustCommand{\orcidicon}{%
	\begin{tikzpicture}
	\draw[lime, fill=lime] (0,0) 
	circle [radius=0.16] 
	node[white] {{\fontfamily{qag}\selectfont \tiny ID}};
	\draw[white, fill=white] (-0.0625,0.095) 
	circle [radius=0.007];
	\end{tikzpicture}
	\hspace{-2mm}
}

\foreach \x in {A, ..., Z}{%
	\expandafter\xdef\csname orcid\x\endcsname{\noexpand\href{https://orcid.org/\csname orcidauthor\x\endcsname}{\noexpand\orcidicon}}
}

\title[\xmm{} - \erosita{} cross calibration]{The \xmm{} Cluster Survey: Cross-Calibration of \erosita{} and \xmm{} using Dark Energy Survey Y3 Galaxy Clusters}

\author[J. E. Pilling et al.]{J. E. Pilling,$^{1,2}$\thanks{E-mail: j.pilling@sussex.ac.uk (JEP)}\orcidA{},
D. J. Turner$^{3,4,5}$\orcidB{},
P. A. Giles$^{1}$\orcidC{},
A. K. Romer$^{1}$\orcidD{},
R. Wilkinson$^{1}$\orcidE{},
\newauthor
P. J. Rooney$^{1}$,
M. Hilton$^{6,7}$\orcidF{},
S. Bhargava$^{8}$\orcidI{},
J. Hall$^{9}$\orcidH{},
R. Wang$^{10}$\orcidG{},
A. Abdulghafour$^{1}$\orcidJ{},
\newauthor
B. J. Maughan$^{9}$\orcidK{} \\
$^{1}$Department of Physics and Astronomy, University of Sussex, Brighton, BN1 9QH, UK\\
$^{2}$Universität Innsbruck, Institut für Astro- und Teilchenphysik, Technikerstr. 25/8, 6020 Innsbruck, Austria\\
$^{3}$Center for Space Science and Technology, University of Maryland, Baltimore County (UMBC), Baltimore, MD 21250, USA\\
$^{4}$High Energy Astrophysics Science Archive Research Center (HEASARC), NASA/GSFC, Greenbelt, MD 20771, USA\\
$^{5}$Center for Research and Exploration in Space Science and Technology (CRESST II), NASA/GSFC, Greenbelt, MD 20771, USA\\
$^{6}$Wits Centre for Astrophysics, School of Physics, University of the Witwatersrand, Private Bag 3, 2050, Johannesburg, South Africa\\
$^{7}$School of Mathematics, Statistics \& Computer Science, University of KwaZulu-Natal, Westville Campus, Durban 4041, South Africa\\
$^{8}$Université Côte d’Azur, Observatoire de la Côte d’Azur, CNRS, Laboratoire Lagrange, Bd de l’Observatoire, CS 34229, 06304 Nice cedex 4, France\\
$^{9}$School of Physics, HH Wills Physics Laboratory, University of Bristol, Tyndall Avenue, Bristol, BS8 1TL, UK\\
$^{10}$Michigan State University, Physics \& Astronomy Department, East Lansing, Michigan, USA\\
}

\date{Accepted XXX. Received YYY; in original form ZZZ}

\pubyear{\the\year{}}

\begin{document}
\label{firstpage}
\pagerange{\pageref{firstpage}--\pageref{lastpage}}
\maketitle


\begin{abstract}
We present the cross-calibration of the \erosita{} and \xmm{} telescopes, using galaxy clusters detected by redMaPPer from the first three years of Dark Energy Survey observations (DESY3RM). We use the X-ray: Generate and Analyse (\xga{}) Python module to measure \erosita{} properties, and validate our methodology using a deep pointed \erosita{} CAL-PV observation of \democluster{}, and a sample of 119 clusters from the expanded HIgh FLux Galaxy Cluster Sample (eeHiFLUGCS). Next, we construct a sample of DESY3RM clusters matched to \xmm{} observations analysed by the \xmm{} Cluster Survey (XCS), and to clusters in the first release of the eROSITA All-Sky Survey (\erass{}). This produced a sample of 114 DESY3RM clusters with both \xmm{} and \erosita{} detections.  Our resulting X-ray temperatures determined from \erosita{} were compared to \erass{} catalogue values, showing consistency between the two analyses. Comparisons between \xmm{} and \erosita{} temperatures reveal a systematic offset, with \xmm{} temperatures $28\pm7$\% higher on average. Finally, we fit $L_{\rm X,soft}$-$\lambda_{\rm RM}$ and $T_{\rm X}$-$\lambda_{\rm RM}$ scaling relations for both \erosita{} and \xmm{} using the same cluster sample, finding minimal differences in the $L_{\rm X,soft}$-$\lambda_{\rm RM}$ relation but a clear normalisation offset in the $T_{\rm X}$-$\lambda_{\rm RM}$ relation, particularly for the highest richness systems. This study will enable future work combining \erosita{} and \xmm{} observatories for DESY3RM clusters.

\end{abstract}





\section{Introduction}
\label{sec:intro}
Galaxy clusters are tracers of the most massive dark matter halos in the universe. The mass function of dark matter halos across cosmic time is sensitive to cosmological parameters, hence galaxy cluster properties e.g., the cluster mass, are invaluable for cluster abundance cosmology experiments. Such experiments require well-characterised observables that correlate with cluster mass through scaling relations, with weak gravitational lensing commonly used to calibrate the connection between these mass proxies and the underlying halo mass.

The choice of observable to construct a scaling relation varies between experiments, which span a wide range of wavelengths. Galaxy clusters can be studied via: the hot intracluster medium (ICM), using X-ray and millimetre-wavelength observations, their constituent member galaxies via optical observations, and the underlying dark matter component can be probed by gravitational (weak and strong) lensing. The Dark Energy Survey (DES) is a photometric survey conducted in the g, r, i, z and Y bands covering \textasciitilde 5000 square degrees of the sky, using the 570 megapixel Dark Energy Camera \citep[DECam,][]{DECcam2015}. DES uses cluster richness (typically denoted $\lambda$) as the observable, a probabilistic measure of the number of member galaxies, estimated by the red-sequence Matched-filter Probabilistic Percolation cluster \citep[hereafter redMaPPer,][]{redmapper2016} to construct a scaling relation calibrated to masses derived via weak lensing \citep{DESMassCalib2019}. 

The cosmological constraints from the first year of DES observations (DESY1) using cluster abundances and weak lensing \citep[][hereafter A20]{DESY1Clusters} were in tension with other DESY1 analyses and external cosmological experiments. In the $\Omega_{M}$ - $\sigma_{8}$ plane, their results were at a $2.4\sigma$ tension with DES Y1 3x2pt analysis \citep{DESY1Cos3X2pt2018}, and $5.6\sigma$ tension with the Planck Cosmic Microwave Background results \citep{Planck2016}.  As A20 discuss, this tension could be due to the slope of the mass-richness relation being biased high due to an incorrect interpretation of the stacked weak lensing signal for low richness ($\lambda < 30$) clusters. When these clusters were removed from the analysis, the internal tension with DES Y1 $3\times2$pt results is no longer present. Furthermore, A20 uses a stacked weak lensing analysis, thereby losing information of the intrinsic scatter of the scaling relation. \citet{Sahlen2009} demonstrated that the scatter, together with its redshift dependence, needs to be effectively exploited to infer cosmological parameters.

A20 demonstrated the need for multi-wavelength data to better understand systematic effects in their cosmology analysis, primarily via X-ray data from the {\em XMM-Newton}\footnote{\href{https://www.cosmos.esa.int/web/xmm-newton}{https://www.cosmos.esa.int/web/xmm-newton}}(hereafter \xmm{}) and \chandra{}\footnote{\href{https://www.nasa.gov/mission/chandra-x-ray-observatory/}{https://www.nasa.gov/mission/chandra-x-ray-observatory/}} observatories. The X-ray data was used to understand miscentering effects and better describe scaling relations. \cite{zhangDES19} provided a model description of the miscentering of the redMaPPer cluster position compared to the X-ray centroids.  redMaPPer richness scaling relations with respect to X-ray luminosity and temperature are investigated in e.g., \cite{FarahiDES19}, \cite{Hollowood19}, \cite{Giles2022SDSS}, \cite{Kelly2024}. These studies are all similarly limited by the complex selection function of the \xmm{} and \chandra{} observing strategies, as they are largely driven by follow up of known objects, rather than serendipitous discoveries.  Furthermore, studies using \xmm{} contiguous regions are limited by the size of the survey overlap with DES, comprising only $\sim$60 deg$^{2}$ \citep{Upsdell2023}. 

The \erosita{} X-ray telescope \citep{erosita} on board the Spectrum Roentgen Gamma observatory launched in 2019. With the public release of the first six months of the \erosita{}-All Sky Survey within the western galactic hemisphere (known as \erass{}), $20626.5$\,deg$^{2}$ of X-ray data became available \citep{erasssrccat}. This offers a unique opportunity to support optically selected cosmological samples with X-ray data that has a well-characterised selection function. Aside from DES, forthcoming cluster abundance cosmology analyses from surveys such as Euclid \citep{euclid} and LSST \citep{LSST} require an understanding of the systematic effects associated with scaling relations. This work aims to contribute to this effort.

We revisit the X-ray support provided by \xmm{} to DES, by comparing measurements of ICM temperature and luminosity with those measured by \erosita{}. Systematic offsets of physical properties of galaxy clusters derived from different X-ray observatories have been well studied. This has been characterised by comparing \xmm{} and \chandra{} cluster temperatures in e.g., \cite{Nevalainen2010}, \cite{Schellenberger2015}. The work of \cite{Lopez2025} performed a cross-calibration with NuSTAR \citep{nustar}, \xmm{}, and \chandra{}. The first \erosita{} cross-calibration \citep[][hereafter \citetalias{Turner2022}]{Turner2022} found \xmm{} temperatures $25 \pm 9 \%$ larger than \erosita{}.  Following the release of eRASS1, \cite{Migkas2024} studied \erosita{} cross-calibrations, finding a $14\%$ offset with \xmm{} (with a small dependence on cluster temperature). While not focused on cross-calibration, \cite{Kelly2024} found that optical to X-ray scaling relations are affected by the choice of X-ray observatory. It was found that the slope of the richness - X-ray temperature relation differs depending on the use of \chandra{} or \xmm{} for X-ray analysis. It is therefore valuable to characterise this calibration between X-ray instruments to enable future studies to collate data from multiple observatories, and to understand the effect on scaling relations that support cluster abundance cosmology studies.


The work presented herein represents the first application of the X-ray: Generate and Analyse \citep[\xga{},][]{xga2022,asclxga} software package\footnote{\label{foot:xga}\href{https://github.com/DavidT3/XGA}{https://github.com/DavidT3/XGA}} to \erosita{} data. Previously, \xga{} could only make use of \xmm{} observations \citep[e.g.][]{turner2025}, but a new release adds support for \erosita{}. 
We demonstrate the unique capability of \xga{} to measure X-ray properties of galaxy cluster populations from both \erosita{} and \xmm{} using a consistent methodology. Furthermore, we describe the \xga{} procedure for deriving higher-level properties, such as entropy and hydrostatic mass profiles, using the example of \democlusterfull{} (hereafter \democluster{}), which has pointed observations from both \xmm{} and \erosita{}.
These results allow us to establish an \erosita{}–\xmm{} cross-calibration applicable to future DES X-ray support studies employing both observatories.

The outline of this paper is as follows; Section~\ref{sec:data} details the preparation of data from the \erosita{} and \xmm{} archives, ready for \xga{} ingestion. Section~\ref{sec:sample} introduces the sample of objects that have been detected by redMaPPer, the \xmm{} Cluster Survey \citep{XCS2001}, and \erosita{}. In Section~\ref{sec:method} we present the updated X-ray: Generate and Analyse (\xga{}) software package, now including \erosita{} compatibility, and apply it to \democluster{} as a demonstration of the methodology. We validate the \xga{} method in Section~\ref{sec:valid}, by performing comparisons of X-ray temperatures and luminosities ($T_{\rm X}$\footnote{Unless otherwise specified, we use $T_{\rm X}$ to denote a global value measured within \rad{}, where \rad{} represents a radius at which the density of the cluster becomes 500 times that of the critical density of the University at the cluster redshift.} and $L_{\rm X, soft}$\footnote{``soft'' refers to $L_{\rm X}$ values measured across the 0.5 - 2.0 keV band})  measured by \xga{} to the \erass{} cluster catalogue \citep[][hereafter \citetalias{Bulbul2024}]{Bulbul2024}, and annular temperatures ($T_{\rm X}^{CE}$\footnote{CE = core excluded, from 0.2 - 0.5 \rad{}}) to clusters in \citet[][hereafter \citetalias{Migkas2024}]{Migkas2024} which were selected from the expanded HIgh FLux Galaxy Cluster Sample (eeHIFLUGCS). In Section~\ref{sec:XMMCalib} we provide a $T_{\rm X}$, $L_{\rm X, soft}$, and $L_{\rm X, bol}$\footnote{``bol'' refers to $L_{\rm X}$ values measured across the 0.1 - 10.0 keV band} calibration of \erosita{} to \xmm{} for the DES subsample described in Section~\ref{sec:sample}. We also provide a comparison of optical - X-ray scaling relations as measured by \erosita{} and \xmm{} in Section~\ref{sec:scalingrel}. Finally we discuss origins of offsets between observatories in Section~\ref{sec:Dis} and provide our conclusions in Section~\ref{sec:conc}. For all \xga{} analysis we assume a cosmology of $\Omega_{M}$=0.3, $\Omega_{\Lambda}=0.7$, and $H_{0}= 70$ km s$^{-1}$ Mpc$^{-1}$.

\section{Data}
\label{sec:data}
In all subsequent analyses, we employ the open-source Python package \xga{}. Before \xga{} can be used, it requires a set of science-ready event lists and, optionally, region files describing the spatial extent of detected sources. Sections \ref{subsec:erosita} and \ref{subsec:erositaedr} describe the \erosita{} data from the survey and the calibration and performance verification phases respectively, and their corresponding source regions, which are input into \xga{}. Section \ref{subsubsec:daxa} details the preparation of the \erosita{} data, making use of the open-source Python package \daxa{} \citep{daxa}. Section \ref{subsec:xmmobs} details the equivalent data products for \xmm{}. Finally, Section \ref{subsec:desrm} describes the optical data and cluster detection from DES.

\subsection{\erass{}}
\label{subsec:erosita}
We make use of the public \erass{} data\footnote{\label{foot:erass1}\href{https://erosita.mpe.mpg.de/dr1/erodat/}{https://erosita.mpe.mpg.de/dr1/erodat/}}, a contiguous area covering the entire western galactic hemisphere, observed from December 2019 to June 2020. The observing strategy of \erass{} is a constant slew around a series of great circles converging at the ecliptic poles. The result is an average exposure time of $\sim$100\,s at the ecliptic equator to $\sim$10,000\,s at the ecliptic poles \cite[see][for full details, and in particular Figure~3]{erasssrccat}. The slewing of the telescope means that source photons arrive at the detector across different positions in the field of view, such that sources will be subject to the spatially varying effective area. Consequently, calculations requiring the effective area of the telescope, such as exposure maps and ancillary response files for spectral calibration, must be done in time steps corresponding to the speed of sources across pixels on the detector. This impacts our choices for spectral analysis which are discussed further in Section~\ref{subsec:specxga}.

All \erass{} data processing was done with the \erosita{} Science Analysis Software System for Data Release 1 \citep[hereafter \esass;][]{Brunner2022}\footnote{\label{foot:esass}\href{https://erosita.mpe.mpg.de/dr1/eSASS4DR1/}{https://erosita.mpe.mpg.de/dr1/eSASS4DR1/}}. 


\subsubsection{\erass{} Source Regions}
\label{subsubsec:erasssrc}
Throughout this work, we default to using \xmm{} regions (described in Section~\ref{subsec:xmmobs}) to mask contaminating sources from our analysis. In the cases where \xmm{} data is unavailable (see Section~\ref{subsec:setupxga}), the following procedure is used. For the detection and masking of contaminating sources within \erass{} data, we use the \texttt{EXT} parameter within the \erass{} primary source catalogue \citep[][]{erasssrccat}. \texttt{EXT} represents the core size (in arcseconds) of a beta model fitted to each source detected in the \cite{erasssrccat} pipeline. Using this parameter, we assume that the full spatial extent of each source corresponds to a circular region with a radius given by the `Masking Radii' Column of Table \ref{tab:eromasking}. 


\begin{table}
\begin{center}
\caption[]{Spatial extent of sources in \erass{} data, used to mask contaminating sources within analysis regions. The `Condition' column lists the criteria for masking a source, with the corresponding `Masking Radii' (next column) specifying the circle radius applied. \texttt{EXT} refers to the \texttt{EXT} column in \cite{erasssrccat} (in arcseconds).}\label{tab:eromasking}
\vspace{1mm}
\begin{tabular}{p{0.19\linewidth}|p{0.15\linewidth}|p{0.35\linewidth}}
\hline
\hline
Condition & Masking Radii & Notes \\
\hline
\texttt{EXT} > 0 & 6${\times}$\texttt{EXT} & Assuming every extended source is a cluster. 6${\times}$\texttt{EXT} corresponds to \textasciitilde\rad{}.\\
\hline
\texttt{EXT} = 0 & 30\arcsec & Point source masking set to FoV-averaged resolution. \\
\hline
\hline
\end{tabular}
\end{center}
\end{table}


\subsection{\erosita{} CAL-PV}
\label{subsec:erositaedr}
In this work we make use of the calibration and performance verification phase of \erosita{} (\erosita{} CAL-PV\footnote{\href{https://erosita.mpe.mpg.de/edr/}{https://erosita.mpe.mpg.de/edr/}}). This phase took place between September and December 2019, carrying out pointed observations as well as scan observations, including the \erosita{} Final Equatorial Depth Survey \citep[eFEDS,][]{Brunner2022}. We use the  pointed observation of galaxy cluster \democluster{}, with observation ID 700177. Figure~\ref{fig:a3158regs} shows the 0.5 - 2.0 keV image of this observation, using the full \erosita{} field of view (0.515$^\circ$ radius), with the centre of \democluster{} (R.A., Dec.) = (55.71$^\circ$, -53.63$^\circ$) indicated by a white crosshair. The central coordinates and redshift ($z$ = 0.059) of \democluster{} are taken from \citet[][hereafter \citetalias{Whelan2022}]{Whelan2022}. Since this observation has an exposure time of $\sim$80,000 seconds, it enables high enough signal to noise to provide constraints on analysis products up to the level of hydrostatic mass profiles. Data processing of this observation was done with the early data release version of \esass{}\footnote{\label{foot:esassedr}\href{https://erosita.mpe.mpg.de/edr/DataAnalysis/}{https://erosita.mpe.mpg.de/edr/DataAnalysis/}}.

\subsubsection{\erosita{} CAL-PV Source Regions}
\label{subsubsec:EROCalPvRegs}
\begin{figure}
    \centering
    \includegraphics[width=1.\columnwidth]{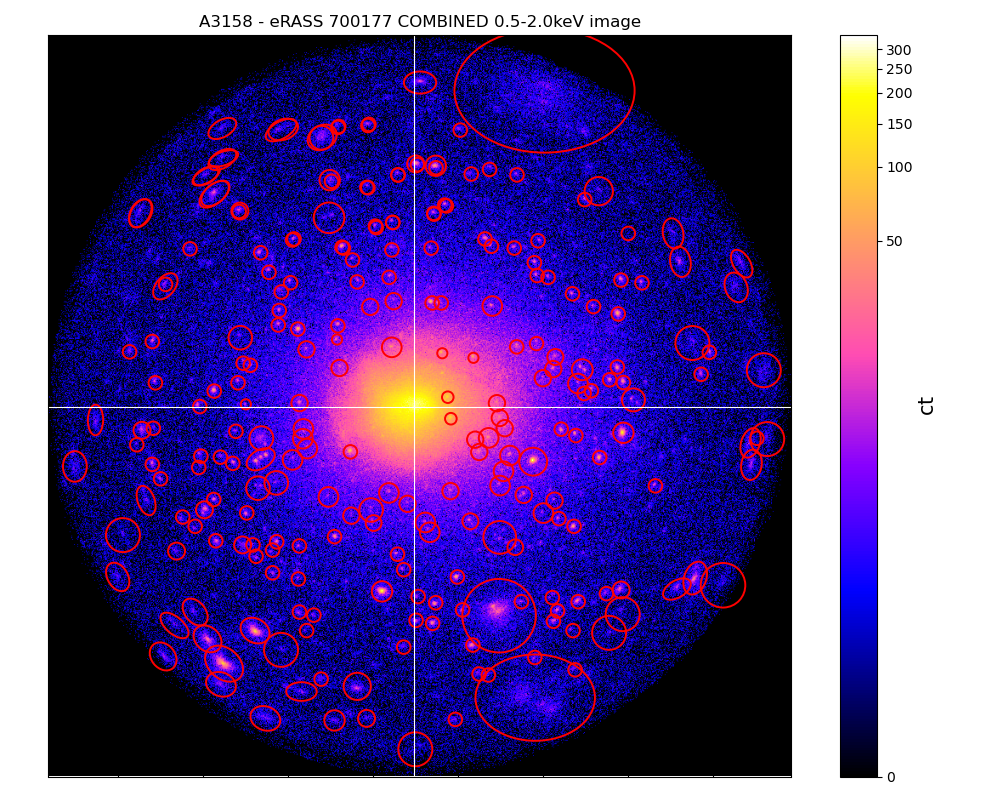}
    \caption[]{The \erosita{} CAL-PV observation (observation ID 700177) of galaxy cluster \democluster{} at $z$ = 0.059. The image shows the full \erosita{} field of view (0.515$^\circ$ radius), with white cross hairs indicating the \democluster{} centre of (R.A., Dec.) = (55.71$^\circ$, -53.63$^\circ$) (taken from \citetalias{Whelan2022}), from which annular spectra are generated. The manually selected contaminating regions, created using the \xga{} interactive region editor tool, are indicated by the red ellipses, and are excluded from all analysis. The contaminating regions for \xmm{} are shown in the right panel of Figure~\ref{fig:splodge_mask}.}
    \label{fig:a3158regs}
\end{figure}

For the observation of \democluster{}, identification of contaminating sources was done manually via the inbuilt \xga{} region editor described in \cite{turner2025}. This differs from the method described in Table \ref{tab:eromasking} as there were no point source catalogues available for this observation. The manually identified contaminating sources are excluded from subsequent analysis, and are shown by red ellipse regions in Figure~\ref{fig:a3158regs}. These masks can be found in the accompanying GitHub repository for this study\footnote{\href{https://github.com/jessicapilling/eRO-XMM_Cross_Calibration}{https://github.com/jessicapilling/eRO-XMM\_Cross\_Calibration}}.

\subsection{Acquiring and Processing \erosita{} Data with \daxa{}}
\label{subsubsec:daxa}
The open-source Python module Democratising Archival X-ray Astronomy\footnote{\label{foot:daxa}\href{https://github.com/DavidT3/DAXA}{https://github.com/DavidT3/DAXA}} \citep[hereafter \daxa{};][]{daxa} facilitates the full exploitation of archival data from X-ray missions including \erosita{}, \xmm{}, \chandra{}, \textit{Swift} \citep{swift}, \textit{NuSTAR} \citep{nustar}, and \textit{XRISM} \citep{xrism}. \daxa{} allows users to interact with data archives of various X-ray telescopes in a consistent manner through Python; acquire said archives; conduct the cleaning and pre-processing steps; and update those archives with new observations. 

We obtain \erosita{} data using \daxa, and employ the built-in \daxa{} function, {\tt full\textunderscore process\textunderscore erosita}, for processing. This function internally calls the \esass{} {\tt FLAREGTI} command, with parameters set to the recommended defaults. {\tt FLAREGTI} appends Good Time Intervals (GTIs) to the unprocessed event lists, excluding time periods affected by soft-proton flaring. These GTIs are then applied using the \esass{} command {\tt EVTOOL}, which retains only valid single-, double-, triple-, and quadruple-pixel events, while removing those flagged as corrupt or belonging to corrupt frames.

\subsection{\em{XMM-Newton}}
\label{subsec:xmmobs}
For \xmm{} analysis we use public data from the \xmm{} Science Archive\footnote{\href{http://nxsa.esac.esa.int/nxsa-web/}{http://nxsa.esac.esa.int/nxsa-web/}}, pre-processed according to the \xmm{} Cluster Survey \citep[XCS;][]{XCS2001} pipeline as described in \cite{Davies2011} and \cite{Giles2022SDSS}, yielding event lists cleaned of time periods with a high background. For \xmm{} contaminating source detection and masking we use the XCS Automated Pipeline Algorithm (\xapa{}), based on {\tt WAVDETECT} \citep{xapa2002}. {\tt WAVDETECT} is described in Section 3.1 of \cite{Davies2011}. For all \xmm{} product generation, the \xmm{} Science Analysis Software (\textsc{SAS}) version 22.1.0 was used.

\subsection{DES Y3 RedMaPPer Cluster Catalogue}
\label{subsec:desrm}
In this work we make use of galaxy clusters identified by redMaPPer applied to the first three years of Dark Energy Survey observations (hereafter DESY3).  The DESY3 redMaPPer catalogue used in this work is presented in \cite{abbott2025}. We use the volume-limited, publicly available catalogue\footnote{\label{foot:redmapper}\href{https://des.ncsa.illinois.edu/releases/y3a2/Y3key-cluster}{https://des.ncsa.illinois.edu/releases/y3a2/Y3key-cluster}}, containing 21,092 galaxy clusters between redshifts of 0.1 $< z <$ 0.93 and a richness range spanning $20 < \lambda < 222$.  The redMaPPer algorithm self-trains the red sequence model to available spectroscopic redshifts and iteratively estimates photometric redshifts for each detected cluster. A calculation of cluster richness (hereafter $\lambda_{RM}$) is performed, which is defined as the sum of membership probabilities of all galaxies within a scale radius, $R_\lambda$, defined by $R_\lambda = 1.0 h^{-1}$Mpc$(\lambda/100)^{0.2}$. Additional information about redMaPPer and examples of its scientific applications can be found in: \citet{Rykoff2014, Rozo2014, Rozo2015a, Rozo2015, Simet2017, DESMassCalib2019, Giles2022SDSS}.

\section{Samples of Galaxy Clusters}
\label{sec:sample}
In this section we describe the galaxy cluster samples analysed in this work. To demonstrate the full suite of \xga{} analysis tools for galaxy clusters, we use the \erosita{} CAL-PV observation of \democluster{}. In the proceeding Section~\ref{sec:method}, we apply the \xga{} methodology to larger samples to validate the method and perform cross-calibrations. The remainder of this section describes those galaxy cluster samples.

\subsection{Validation Samples}
\label{sec:validation_sample}
The following galaxy cluster samples are used for validating the \xga{} method of measuring the X-ray temperature and luminosity from \erosita{} observations.

The first sample is drawn from \citet[][hereafter \citetalias{Migkas2024}]{Migkas2024} and is publicly available from a GitHub repository\footnote{\label{foot:m24data}\href{https://github.com/kmigkas/eROSITA-cross-calibr-data}{https://github.com/kmigkas/eROSITA-cross-calibr-data}}. This sample is a subsample of the expanded HIgh FLux Galaxy Cluster Sample (eeHIFLUGCS) as detailed in \cite{migkas2020}. Briefly, cluster detections from the Meta-Catalogue of X-ray detected Clusters of Galaxies \citep{mcxc} were selected based on a threshold of an unabsorbed 0.1-2.4 keV X-ray flux of $\geq$  $5 \times 10^{-12}$ ergs~s~$^{-1}~\mathrm{cm}^{-2}$.  The \citetalias{Migkas2024} sample comprises 152 clusters in total. For this work, we focus on the subset of \citetalias{Migkas2024} with a measured temperature within an annular radius of [0.2 - 0.5]$R_{500}$ (hereafter $T_{\rm X}^{CE}$) from \erosita{}, yielding the 119 clusters used in Section \ref{sec:valid}.


The second sample is taken from the \erass{} cluster catalogue presented in \citet[][hereafter \citetalias{Bulbul2024}]{Bulbul2024}. In brief, these clusters were detected in the \erass{} data in the 0.2-2.3 keV band using the source detection pipeline in \cite{erasssrccat}. \citetalias{Bulbul2024} applies a selection of \texttt{EXT} greater than 0$^{\prime\prime}$ and an extent likelihood ($\mathcal{L_{\rm ext}}$) greater than 3, followed by a cleaning procedure to remove regions of bright Galactic emission and cases where a single extended source was artificially segmented into multiple detections. An optical confirmation and redshift allocation process was then carried out, as described in \cite{Kluge2024}, resulting in a final catalogue of 12,247 galaxy groups and clusters.

\subsection{DES-XCS-eRO}
\label{subsection:des-xcs-ero}

In this work we apply \xga{} to a subsample of the 12,247 \citetalias{Bulbul2024} clusters (Section~\ref{sec:validation_sample}) that are also  matched to DESY3 redMaPPer catalogue entries covered by \xmm{} observations. First, DESY3 redMaPPer clusters within the \erass{} footprint were identified.  For a representation of the full overlap of DES with the eRASS1, see Fig.~12 in \cite{Upsdell2023}.  Next these DESY3 redMaPPer clusters were matched to \xmm{} observations via a coordinate search. If a given DESY3 redMaPPer cluster centroid position fell within 15$^{\prime}$ of the aimpoint of one or more {\em XMM} observations, then that cluster was flagged as having an {\em XMM} match.  A visual inspection of \erass{} images, DES cutouts, and \xmm{} images was conducted to ensure that the same source was being detected across all three telescopes. For this visualisation, we used the OCTAVIUS (Object Classification Tools for Astronomy VIsUaliSation) platform\footnote{\href{https://furciferi.github.io/OCTAVIUS}{https://furciferi.github.io/OCTAVIUS}}, a dynamic and interactive web interface for visualising astronomy images with the functionality of classification. We ensured redshift consistency between \citetalias{Bulbul2024} and the DES Y3 redMaPPer catalogue, in order to remove false matches due to projection effects. One such case was identified, in which the \citetalias{Bulbul2024} cluster 1eRASS J032951.5–523451 (with $z=0.0599$) was preliminary associated with a redMaPPer system \texttt{mem\_match\_id} = 43189 (with $z_{\lambda}=0.44$). Visual inspection of the galaxies in the DES cutout revealed a foreground and background cluster projected in close proximity, indicating that this was not a genuine physical association. This system was therefore removed from the sample, resulting in a final sample of 114 clusters (we refer to this sample hereafter as the DES-XCS-eRO sample). None of the clusters selected had a \texttt{P\textsubscript{cont}}
 value > 0.5, where \texttt{P\textsubscript{cont}}
 is the probability of the source being a contaminant, as defined by \citetalias{Bulbul2024}.

\begin{figure}
    \centering
    \includegraphics[width=1.\columnwidth]{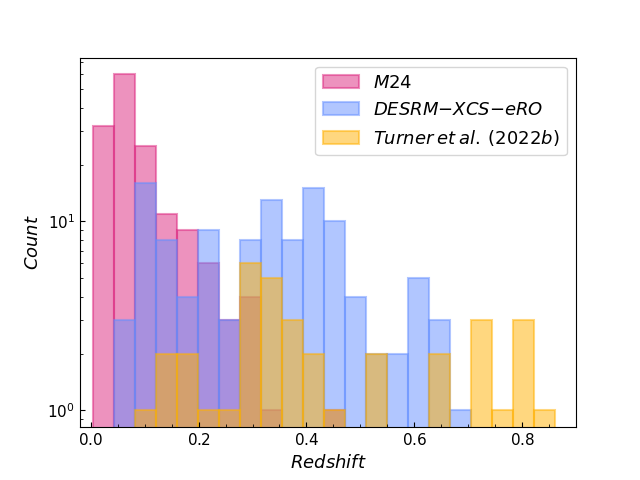}
    \caption{Redshift distribution of the DES-XCS-eRO sample (blue) used in this work for cross-calibration between \erosita{} and \xmm{} (see Section~\ref{subsection:des-xcs-ero}), compared with previous cross-calibration samples: the \citetalias{Migkas2024} sample (pink) and the \protect\citetalias{Turner2022} sample (yellow).}

    \label{fig:redshiftdist}
\end{figure}

We compare the redshift distributions of previous samples used for \erosita{} and \xmm{} galaxy cluster cross-calibration in Figure~\ref{fig:redshiftdist}. The sample used in this work spans a wider redshift range than that of \citetalias{Migkas2024}, and is comparable to the sample of \citetalias{Turner2022}, though it contains a larger number of clusters.

\section{Methodology}
\label{sec:method}

Here we present the cluster analysis methodology using \xga{}. \xga{} is an open-source Python module, listed on the Astrophysics Source Code Library \citep[ASCL;][]{asclxga}, that facilitates X-ray astronomy through data product generation from cleaned event lists and subsequent analysis of those products. It allows the user to interact with and analyse X-ray data consistently within a Python environment, promoting reproducible and transparent science. Previous versions of \xga{} exclusively facilitated \xmm{} analysis. Henceforth, we have developed \xga{} for compatibility with the \erosita{} telescope, to normalise the interface of X-ray analysis across multiple instruments and make archival data exploration accessible and reproducible. In the following subsections we introduce \erosita{} analysis within \xga{} and demonstrate the full suite of galaxy cluster analysis available. We also include the equivalent analysis with \xmm{}, demonstrating the capability of \xga{} to carry out analysis with two telescopes through a consistent interface. In this section we only detail the \xga{} \erosita{} methodology because the \xga{} methodology for \xmm{} was presented in \citet[][hereafter \citetalias{turner2025}]{turner2025}.

\subsection{Setting up \xga{}}
\label{subsec:setupxga}
To ingest galaxy clusters into \xga{}, the coordinates, redshift, and annular radius (e.g., an overdensity radius such as \rad{}) must be provided. For \democluster{} we use (R.A., Dec.) = ($55.71^\circ$, $-53.63^\circ$), $z = 0.059$, and  \rad{} = 1,200 kpc. For the \citetalias{Migkas2024} validation sample, we use the values available in \citetalias{Migkas2024} for the central coordinates, redshifts, and \rad{} values. For the DES-XCS-eRO sample, we adopt the central coordinates derived by \xapa{} (described in Section~\ref{subsec:xmmobs}), and the redshift and \rad{} calculated in \citetalias{Bulbul2024}. 

Once this initial information is provided, \xga{} searches for relevant observations within whatever prepared data sets have been provided (see Sections~\ref{subsubsec:daxa}~\&~\ref{subsec:xmmobs}). For \erass{}, where the survey is divided into tiles, we adopt a search radius of $3.6^\circ$ around each cluster’s central coordinate to ensure that clusters near tile boundaries include all relevant events. For \xmm{} observations, we search within 30\arcmin radius.

Optionally, the user may provide region files to \xga, enabling the automated  masking of contaminating sources. In this work, we supply such files, and the automated masking is utilised for all analysis (see Sections \ref{subsubsec:erasssrc}, \ref{subsubsec:EROCalPvRegs}, and \ref{subsec:xmmobs}). Where \xmm{} data is available, we provide \xapa{}-identified regions to \xga, which are applied to both the \erosita{} and \xmm{} analyses. In cases where no \xmm{} data is available, for example some \erosita{} measured clusters in \citetalias{Migkas2024}, and at larger radii that extend beyond the \xmm{} field of view, we default to using source regions described by Table \ref{tab:eromasking}. \xga{} cross-matches the regions defined in these files with the user-specified analysis coordinates, identifying the nearest extended source as the galaxy cluster region. All other regions are treated as contaminants, and their spatial extent is masked from subsequent \xga{} analysis.

\subsection{Generating Images and Exposure Maps}
\label{subsec:imgexpxga}

In this section we focus on the methodology of generating \erosita{} images and exposure maps with \xga{}.

To account for clusters that span multiple tiles of \erass{}, \xga{} merges relevant event lists using the \esass{} {\tt EVTOOL} command, producing a combined event list from which subsequent data products are generated. For image generation \xga{} again calls {\tt EVTOOL}, generating images in the 0.5 - 2 keV band. Each image pixel is set to be 4.35\arcsec, matching the pixel size adopted by \xapa{} for \xmm{} images (i.e. the default pixel size used by XCS, \cite{Davies2011}, Section \ref{subsec:xmmobs}). Exposure maps are produced using the {\tt EXPMAP} command with the parameters {\tt withweights=yes}, to weight each of the seven telescope modules by their relative contributions to the effective area, and {\tt withdetmaps=yes}, to exclude exposure contributions from the edges of the detector where the effective area is challenging to characterise reliably. Once images and exposure maps of equivalent energy bands have been generated, \xga{} automatically computes a ratemap.

\subsection{Generating and Fitting Spectra}
\label{subsec:specxga}
Spectral extraction for \erosita{} is performed on a merged event list, which is the combination of all \xga{}-identified relevant observations (as described in Section~\ref{subsec:imgexpxga}). \xga{} interfaces with the \esass{} command {\tt SRCTOOL} to generate \erosita{} spectra. Due to the \erass{} observing strategy described in Section~\ref{subsec:erosita}, for spectral calibration, the effective area that a source extraction region is subject to varies across time. Since the clusters used in this work have large spatial extents with mixed morphologies, we provide {\tt SRCTOOL} with the image generated in Section~\ref{subsec:imgexpxga} to the {\tt extmap} argument, and set {\tt psftype=NONE}. This allows {\tt SRCTOOL} to generate an ancillary response file that is customised to the source's movement across the telescope's field of view. We increase the {\tt tstep} argument from the default to 0.5, as a compromise between processing time and spectral calibration accuracy. Lastly we use the HEASoft\footnote{\label{foot:heasoft}\href{https://heasarc.gsfc.nasa.gov/lheasoft/}{https://heasarc.gsfc.nasa.gov/lheasoft/}} {\tt ftgrouppha} command to group the output spectrum by a minimum count of 5, in order to improve the signal to noise of each point on the spectrum. In this work we generate a combined spectrum for all seven telescope modules.

For the treatment of the background, \xga{} uses a local annulus to extract a background spectrum, which is then subtracted from the source spectrum after a scaling proportional to the source extraction area is applied. We note that in cases such as the right panel of Figure~\ref{fig:bins}, where the background region extends beyond the instrument field of view, only the area intersecting the detector is used to calculate this scaling.

Due to the clusters covering a large area in the sky, the background annulus spectrum is generated in a separate {\tt SRCTOOL} command run, so that the GTIs from the source region spectra are not used on the background one thereby increasing the signal to noise in the background spectrum. We select the local area for the background by using an annulus centred on the cluster centre with an inner radii of 1.05\rad{}, and the outer radii of 1.5\rad{}.

In this work we perform several  temperature ($T_{\rm X}$) and luminosity ($L_{\rm X}$) measurements. Throughout, spectra are fit within \xga{} using version 12.14.1 of {\sc Xspec} \citep{xspec}. Spectra are fit with an absorbed plasma emission model \citep[\texttt{constant${\times}$tbabs${\times}$apec;}][]{tbabs, apec}. Several parameters are frozen during fitting: the $\rm{n_H}$ value required by \texttt{tbabs} is set using the HEASoft \texttt{nh} command, with the column density taken from \cite{nh}; the metal abundance is fixed at $Z=0.3Z_\odot$; and the redshift is frozen at the respective catalogue value. All fits are performed using the $c$-statistic \citep[][]{cash} over the 0.3 - 7.9 keV range, and with the abundance table of \cite{aspl}. 

The fitting method described above is used by \xga{} for spectra extracted from cluster-wide, core excluded, and annular regions.  For the cluster-wide (or ``global'') spectra, a circular region centred on input centroid was used. The radius of this region can be set in \xga{}. For example \rad{} was used in Sections~\ref{subssec:primcat} and \ref{sec:XMMCalib}. For core excluded spectra, a single circular annulus was used. The inner and outer radii can be set in \xga{}. For example $0.2<0.5$\rad{} was used in Section~\ref{subsec:M24}. For annular spectra, the regions were selected as described below (Section~\ref{subsubsec:calcannuli}).




\subsection{Analysis of Radial Profiles}
\label{subsec:profxga}
In the following section we provide summaries of the \xga{} methodology to generate profiles of: temperature, gas density, hydrostatic mass and entropy. For a comprehensive description see \citetalias{turner2025} (and specifically Section 2 of that paper). We note that in the latest version of \xga{}, functionality has been added to measure entropy profiles, which is described in full in Section~\ref{subsubssec:entropyprofs}.

\subsubsection{Calculating Annuli Widths for Spectral Analysis}
\label{subsubsec:calcannuli}
\begin{figure*}
  \centering
  \begin{subfigure}{0.45\textwidth}
    \centering    \includegraphics[width=\linewidth]{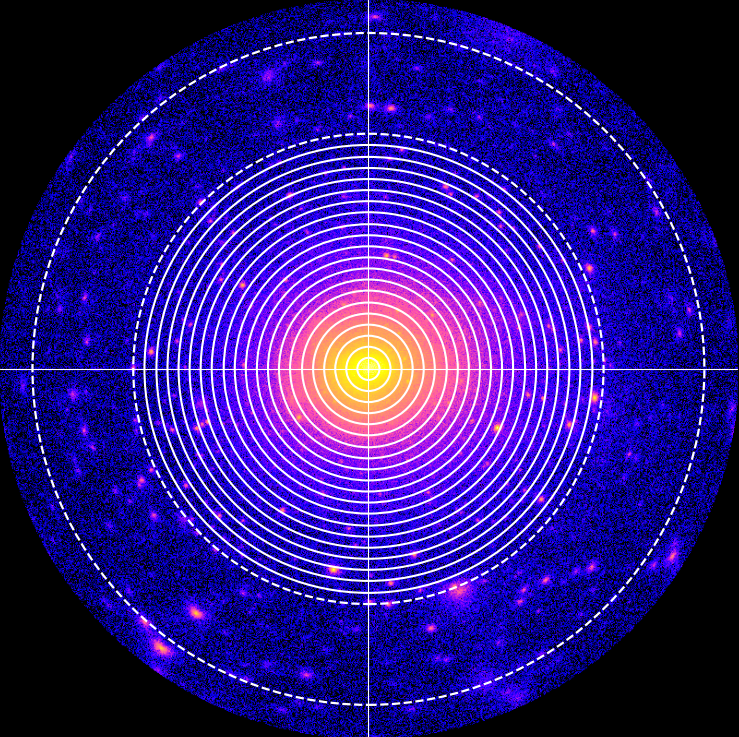}
    \label{fig:subBinsEro}
  \end{subfigure}
  \hfill
  \begin{subfigure}{0.45\textwidth}
    \centering
    \includegraphics[width=\linewidth]{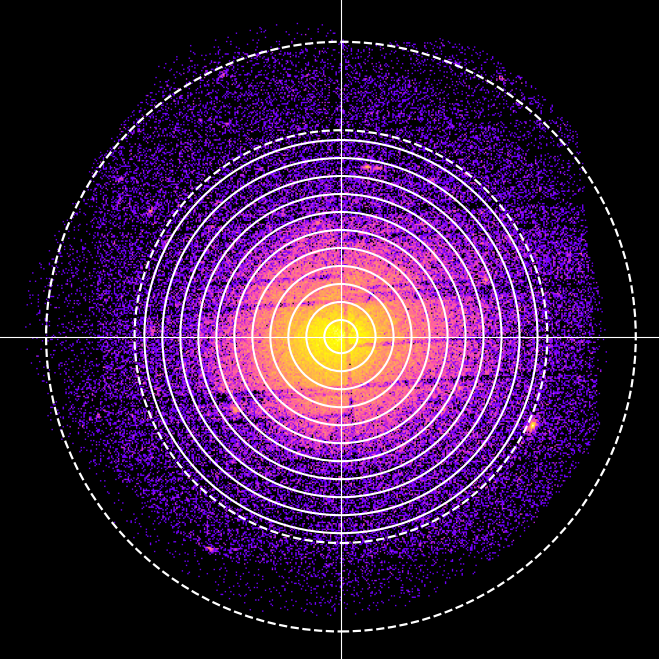}
    \label{fig:subBinsXmm}
  \end{subfigure}
  \caption{X-ray count rate maps in the 0.5 - 2.0 keV band of \democluster{} made with \erosita{} (Left) and  \xmm{} (Right). The annular regions used to generate radial profiles (see Figure~\ref{fig:projtempprof}) are shown with solid white lines. The outer, background subtraction, annulus, is shown with dashed lines.}
  \label{fig:bins}
\end{figure*}

\begin{figure*}
  \centering
  \begin{subfigure}{0.48\textwidth}
    \centering
    \includegraphics[width=\linewidth,trim={1.5cm 0.8cm 2.5cm 0.8cm}, clip]{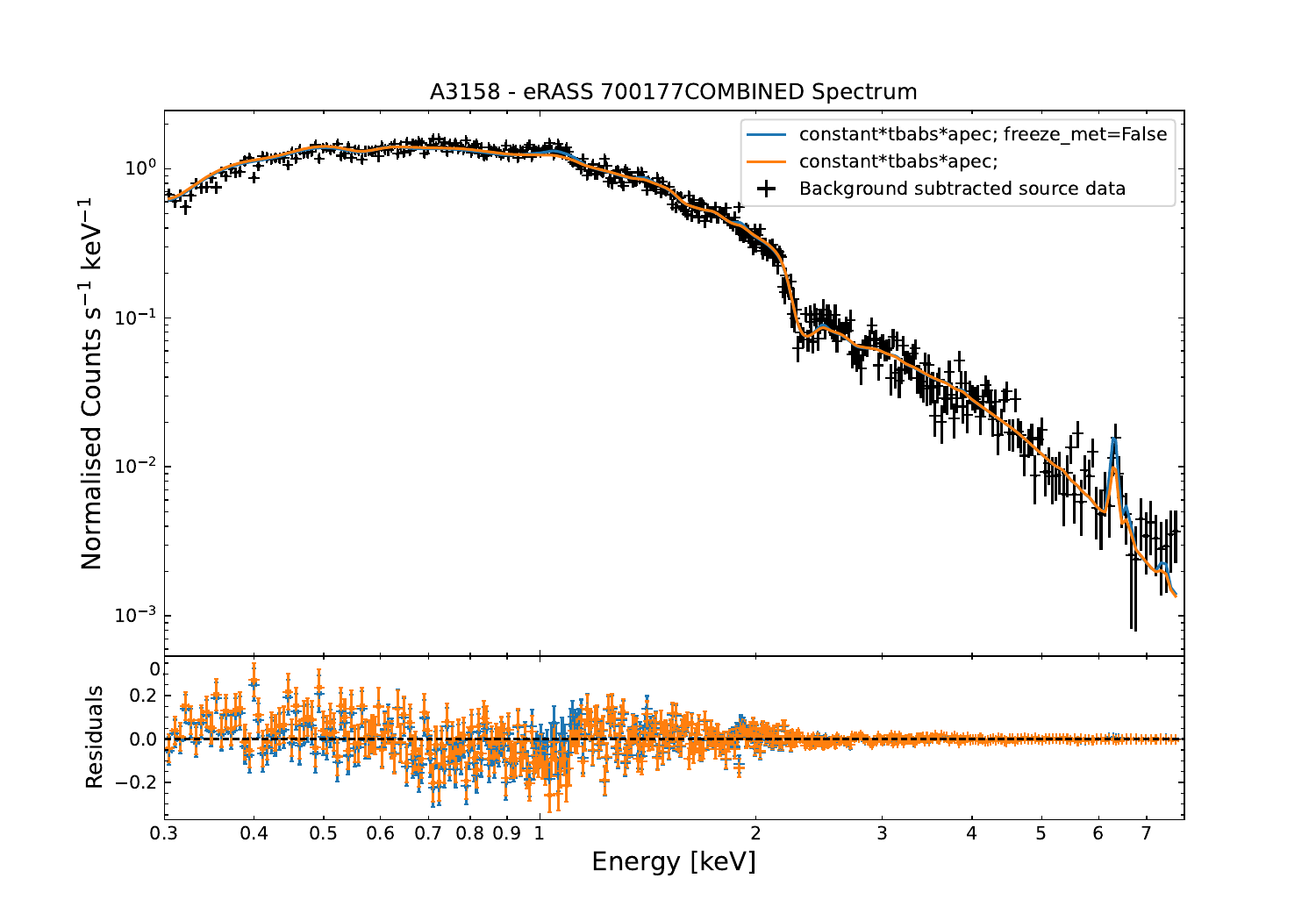}
    \label{fig:subEroSpec}
  \end{subfigure}
  \hfill
  \begin{subfigure}{0.48\textwidth}
    \centering
    \includegraphics[width=\linewidth,trim={1.5cm 0.8cm 2.5cm 0.8cm}, clip]{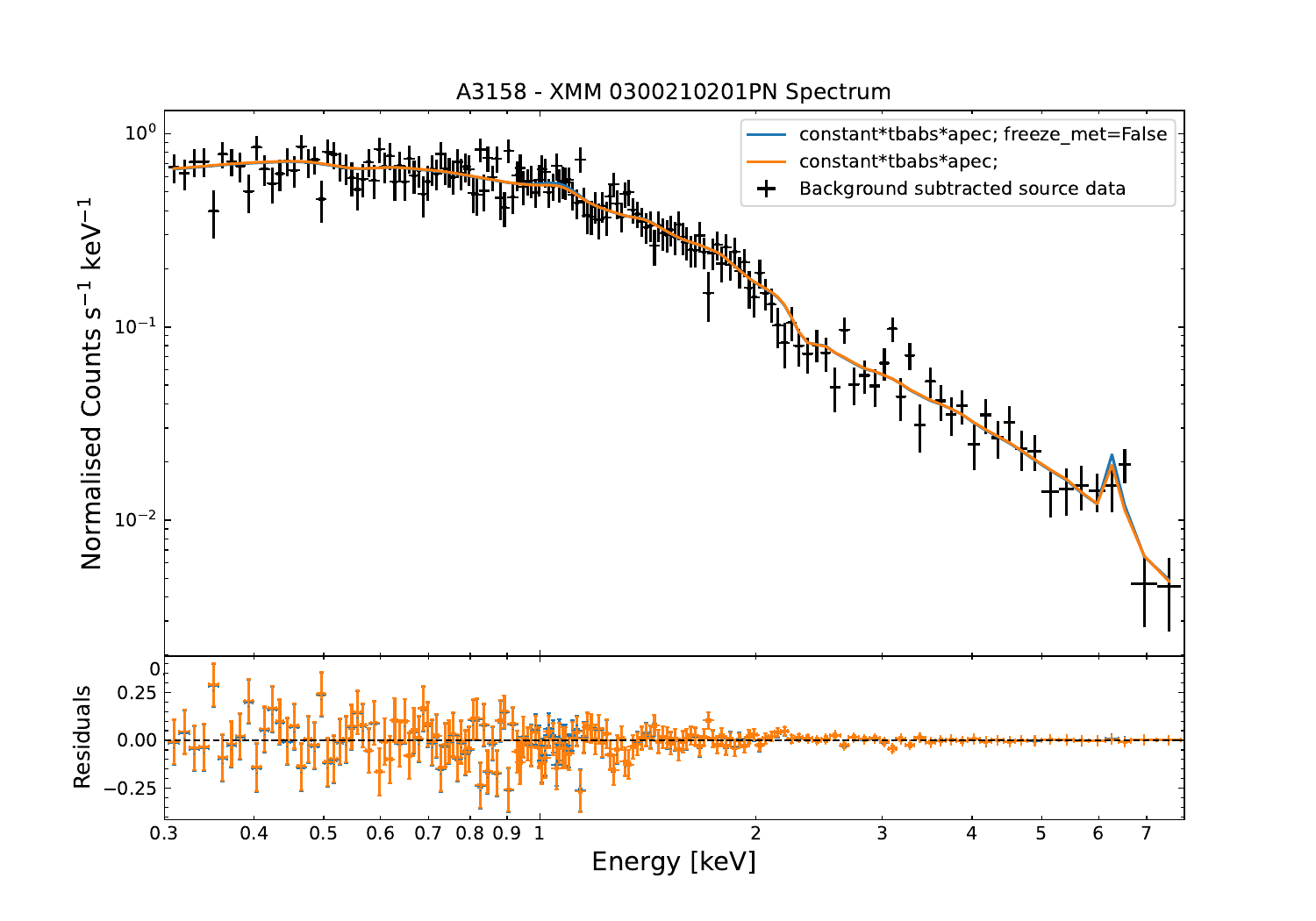}
    \label{fig:subXmmSpec}
  \end{subfigure}
  \caption{Example \xga{} generated \democluster{} spectra from \erosita{} (Left) and \xmm{} (Right). The spectra were extracted from the 0.20 - 0.26 \rad{} annulus. The \xga{} fitting to a  
\texttt{constant$\times$tbabs$\times$apec} model with metallicity ($Z$) and Hydrogen column density ($n_H$) frozen is shown with the orange line and residuals. Also with $n_H$ frozen and $Z$ free (blue line/residuals).  For the purposes of this illustration, a  minimum count (or ``grouping'') of 100 was used (for the rest of the analysis in this paper, 5 was used). Left: \erosita{} combined spectrum from all seven telescope modules. Right: \xmm{} spectrum from the PN instrument.}
  \label{fig:fittedspec}
\end{figure*}

For all profiles requiring spectral analysis, a series of concentric circular annuli centred on the central coordinates are constructed.  See Figure~\ref{fig:bins} for the visualisation of the annuli used during the analysis of \democluster{} and Figure~\ref{fig:fittedspec} for examples of fitted annular spectra. The choices of radii for the annuli were the following:
\begin{itemize}[leftmargin=*, labelindent=2pt, itemsep=1ex]
    \item The outermost radius of the last annuli is set to 1,200 kpc for \erosita{}, and 700 kpc for \xmm{}.
    \item The minimum width of an annulus ($\theta_{\rm min}$) is set to 50\arcsec so that individual spectra have enough signal to noise to produce a reliable fit, and to encompass an area larger than the point spread functions of both telescopes.
    \item We impose a minimum of 1,000 background subtracted counts in each annulus, where the counts are determined from an image generated in the energy band of 0.5 keV to 2.0 keV.
    \item The outermost radius, $\theta'_{outer}$, is adjusted outwardly, such that an integer number of $\theta_{min}$ are encompassed, preserving the minimum annulus width requirement.
    \item Background subtracted counts within bins of width ($\theta_{min}$) are then calculated. The width of bins is then increased until the minimum count is reached. (The profile fitting does not proceed if three or fewer annular bins are defined.)    
\end{itemize}

\subsubsection{Projected Temperature Profiles}
\label{subsubssec:tempprofs}
\begin{figure}
\includegraphics[width=\linewidth]{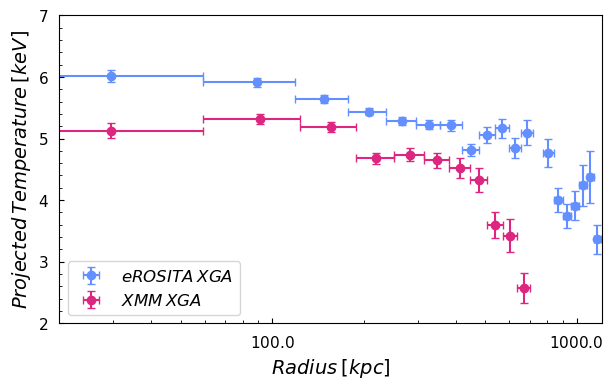}
 \caption{Projected temperature profile of \democluster{} measured with \xga{} for \xmm{} (pink) and \erosita{} (blue). Spectra were extracted from the annular bins shown in Figure~\ref{fig:bins}.}
 \label{fig:projtempprof}
\end{figure}

\xga{} supports the measurement of projected temperature profiles, as well as the deprojection of those measurements to infer the underlying 3D temperature profiles. For projected temperature profiles, the annuli for spectral extraction are determined via the method described in Section~\ref{subsubsec:calcannuli}. For each annulus, spectra are extracted and fitted as described in Section~\ref{subsec:specxga}, thereby obtaining a measurement of temperature. Figure~\ref{fig:projtempprof} shows an \xga{} measured temperature profile for \democluster{} using both \erosita{} and \xmm{} data.

Figure~\ref{fig:projtempprof} demonstrates that, because of its larger field of view, \erosita{} can trace the  \democluster{} projected temperature profiles out to larger radii than \xmm{}. It also shows 
that there is a systematic offset between the profiles at all radii.
This was to be expected: other authors, including using independent (of \xga{}) techniques, have reported offsets between projected $T_X$ values measured by the two telescopes \citepalias{Turner2022, Migkas2024}. However, the direction of the offset (i.e. that \erosita{} is reporting higher $T_X^{\rm Ann}$ values than \xmm{}) is unusual, since previous analyses (and that presented below Section~\ref{sec:XMMCalib}) have found the opposite trend. To ascertain whether this behaviour is intrinsic to \democluster{}, or an artefact of the \xga{} processing, we have compared our analysis to that of \citetalias{Whelan2022}.




\begin{figure}
\includegraphics[width=\linewidth]{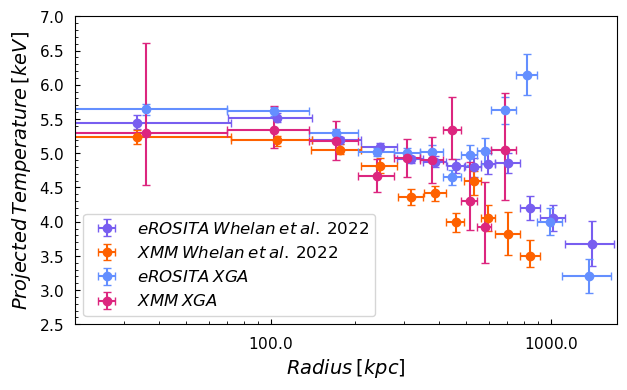}
 \caption{A projected temperature profile of \democluster{}. Similar to Figure~\ref{fig:projtempprof}, but with the \xga{} annuli adjusted to replicate those used by \citetalias{Whelan2022}. The \xga{} fitting methodology has also been modified to match \citetalias{Whelan2022}. The \xga{} results are shown in pink (\xmm{}) and blue (\erosita{}), the  values published in \protect\citetalias{Whelan2022} are shown in orange (\xmm{}) and purple (\erosita{}).}
 \label{fig:projtempprofw22}
\end{figure}

\citetalias{Whelan2022} performed a comparison of \erosita{} to \xmm{} projected temperature profiles using the same \erosita{} CAL-PV observation of \democluster{} shown in Figure~\ref{fig:bins} (left). Figure~\ref{fig:projtempprofw22} shows their results (\erosita{} given by the purple points and \xmm{} by the orange points) and demonstrates that they found a systematic offset the same direction as that seen in  Figure~\ref{fig:projtempprof}. Also shown in Figure~\ref{fig:projtempprofw22} are the results of a second \xga{} analysis of \democluster{} where the \xga{} settings have been modified to more closely follow that of \citetalias{Whelan2022}: we matched their annular bin widths, we allowed the metallicity to vary during the fitting, we set the hydrogen column density to the value they used (1.4$\times 10^{20} \rm cm^{-2}$), we set the same minimum counts (25) per channel when binning spectra, and we used the 0.8 - 9.0 keV range for spectral fitting.  Our $T_X^{\rm Ann}$ results are shown in pink (\xmm{}) and blue (\erosita{}), and it can be seen that in the inner (radius$ < 250$ kpc) bins, the \xga{} and \citetalias{Whelan2022} measurements are in close (within $1 \sigma$) in agreement. However, at larger radii, some of the \xga{} $T_X^{\rm Ann}$ measurements differ significantly from those in \citetalias{Whelan2022}.

 These discrepancies are likely due to the fact that we were not able to exactly replicate the \citetalias{Whelan2022} approach, e.g. \xga{} uses infield, rather than modelled, background subtraction, and our masking decisions (Figure~\ref{fig:a3158regs}) differ. Moreover, \citetalias{Whelan2022} extend to 0.5 keV for the cameras with an on-chip filter, and fit each instrument separately (by comparison \xga{} fits a single combined spectrum). 
 The offsets are most prominent at radii of 600 - 900 kpc. This coincides with a reported region of cool gas (see Figure 14 of \citetalias{Whelan2022}). Therefore, we have applied a additional mask to test the impact of this posited cool gas, finding the \xga{} results to be in better agreement with \citetalias{Whelan2022} as a result. The new mask and the derived projected temperature profile are shown in Appendix~\ref{app:splodge}.



\subsubsection{De-projected Temperature Profiles}
\label{sec:method:deproj}

The temperature profile is de-projected via the `onion-peeling' method described in \cite{onion2002} and \cite{onion2018}. In brief, intersections of 3D shells with projected 2D annuli are calculated using matrix calculations detailed in Appendix A of \cite{onion1999}. This provides a weighting matrix, $V_{\rm int}$, that quantifies the contribution of each annulus to a given 3D shell. By combining these weightings with observed profiles of temperature and normalisation of the {\tt APEC} model (obtained from the spectral fits), their intrinsic 3D distribution can be derived. Uncertainties are obtained by generating 10,000 instances of the projected profiles, thereby calculating 10,000 instances of the 3D profiles. 90\% confidence limits can therefore be obtained at each radial point. 

The upper left panel of Figure~\ref{fig:profs} shows the resulting deprojected temperature profile for \democluster{}. A systematic offset between the profiles is again seen at all radii. The  $T_X^{\rm Ann}$  values are well constrained for both \xmm{} and \erosita{} in the inner regions of the cluster, but become uncertain beyond \textasciitilde 600 kpc. This may be due the existence of multiphase gas at large radii (see above), whereas \xga{} uses a single-temperature model. 

\subsubsection{Density Profiles}
\label{subsubssec:densprofs}
To measure the gas density, we use the normalisation parameter of the {\tt APEC} model, $N_{\mathrm{APEC}}$, represented by, 

\begin{equation}
    N_{\texttt{APEC}} = \dfrac{10^{-14}}{4\pi(D_{A}(1+z))^2}\int n_e n_p dV,
	\label{eq:apecnorm}
\end{equation}

where $D_{A}$ is the angular diameter distance, $z$ is the redshift, and $n_{e}$ and $n_{p}$ are the electron and proton number densities.

$N_{\mathrm{APEC}}$ is obtained from the fits of \texttt{constant${\times}$tbabs${\times}$apec} to the annular spectra described in Section~\ref{subsubsec:calcannuli}. It is then deprojected from a projected annulus measurement to an inferred 3D spherical shell value, $N_{\mathrm{APEC}}(R)$, using the ``onion peeling'' technique (described in Section~\ref{sec:method:deproj}). Equation~\ref{eq:apecnorm} is then re-arranged to give the gas density, assuming an electron-to-proton ratio, $R_{ep}$, using the abundance table of \cite{aspl}:

\begin{equation}
    n_{p}(R) = \sqrt{\dfrac{4\pi N_{\mathrm{APEC}}(R)(D_{A}(1+z))^2}{R_{ep}10^{-14}}}.
	\label{eq:np}
\end{equation}

We note this method differs slightly from that described in Section 2.8 of \citetalias{turner2025}, which uses surface brightness to infer gas density, but both methods are available in \xga{}. In this case we adopt the spectral method because of the high signal-to-noise of the observations.

The resulting gas density profile measured by \xga{} for \democluster{} is shown in the upper right panel of Figure~\ref{fig:profs}. The profiles are in close agreement, with \erosita{} demonstrating that density continues to drop in the cluster outskirts. It is not unexpected that the density profiles are in agreement despite the systematic offset in the  temperature profiles: $T_X^{Ann}$ values depend on the spectral shape, whereas the density is only dependent on the spectral normalisation. A joint Chandra--\xmm{} analysis of 50 clusters by \cite{Martino2014} also reported closely matching density profiles, despite offsets in the $T_X^{Ann}$ profiles.

\subsubsection{Hydrostatic Mass Profiles}
\label{subsubssec:hydromassprofs}
Under the assumption of spherical symmetry and hydrostatic equilibrium, a hydrostatic mass measurement can be obtained through X-ray observations. This can be expressed as follows,

\begin{equation}
\label{equ:hydromass}
    M(r) = - \frac{kr^2}{\rho(r) \mu m_p G} \bigg[\rho(r)\frac{dT}{dr} + T(r)\frac{d\rho}{dr} \bigg],
\end{equation}

where $M(r)$ is the hydrostatic mass at a radius $r$, $k$ is the Boltzmann constant, $\mu$ is the mean molecular weight, $m_p$ is the atomic mass unit, $G$ is the gravitational constant, $\rho(r)$ is the gas density profile, and $T(r)$ is the deprojected temperature profile. An example of the full derivation of this equation can be found in \cite{Fabricant1980}.

Hydrostatic masses are obtained within \xga{} using the models fitted to gas density and de-projected temperature profiles, as measured in Sections~\ref{subsubssec:densprofs} and \ref{sec:method:deproj} respectively. The models fitted to these measurements (as described in Appendix A of \citetalias{turner2025}) have analytical solutions, thereby allowing a direct measurement of their first derivatives at a given radius ($\frac{d\rho}{dr}$ and $\frac{dT}{dr}$ in Equation~\ref{equ:hydromass}). For \democluster{} we present hydrostatic mass measurements in the lower left panel of Figure~\ref{fig:profs}. The profiles are in close agreement, with \erosita{} demonstrating that the enclosed mass continues to rise in the cluster outskirts. As Equation \ref{equ:hydromass} can be reduced to a pressure gradient, the close agreement between the two methods is likely attributable to the relatively flat deprojected temperature profile (top left panel of Figure~\ref{fig:profs}). The small instrumental offset in temperature is compensated by the combined shapes of the density (top right panel of Figure~\ref{fig:profs}) and temperature profiles, yielding a comparable pressure gradient, and therefore a consistent hydrostatic mass profile. This cross-instrument consistency is, like the density profiles, consistent with the findings of \cite{Martino2014}.


\subsubsection{Entropy Profiles}
\label{subsubssec:entropyprofs}
The specific entropy profiles of clusters provide insight into the heating and cooling processes of the hot ICM, as well as acting as a reliable observational diagnostic of whether a cluster is undergoing a merger, or has a prominent cool-core \citep[][]{acceptI_entropy,acceptII_entropy}.

Since \citetalias{turner2025}, \xga{} has been updated to include functionality to measure galaxy cluster entropy profiles, which can be obtained using either a data-driven or model-driven approach. In this work, we present the first use of \xga{} for entropy measurements, and adopt the model-driven method, described in detail below.

To derive entropy, we make use of the temperature and gas density profiles measured in Sections \ref{sec:method:deproj} and \ref{subsubssec:densprofs} respectively. We fit a simplified Vikhlinin Temperature model \citep{Vikhlinin2006} as used by \cite{Ghirardini2019} to the de-projected temperature measurements, and a King model to the gas density measurements; for each, see Appendix A of \citetalias{turner2025} for a full description of the models and fitting procedure. At a given radius, the corresponding three-dimensional temperature and density are obtained from the fitted models in the form of posterior distributions generated from the 10,000 realisations of the profiles. Each realisation is generated using randomly drawn values from the fitted parameter posterior distributions. Entropy is then calculated as:
\begin{equation}
    S(r) = \frac{T_{\mathrm{3D}} (r)}{\rho_{\mathrm{gas}}^{2/3}(r)},
    \label{equ:entropy}
\end{equation}
where $S(r)$, $T_{\mathrm{3D}}(r)$ and $\rho_{\mathrm{gas}}(r)$ are the entropy, de-projected temperature and gas density respectively, at radius $r$. The median value of the resulting entropy distribution is adopted as the best estimate, and the $1\sigma$ uncertainty is defined by the 68.2\% confidence interval. 

\begin{figure*}
    \centering
    \includegraphics[width=0.45\linewidth]{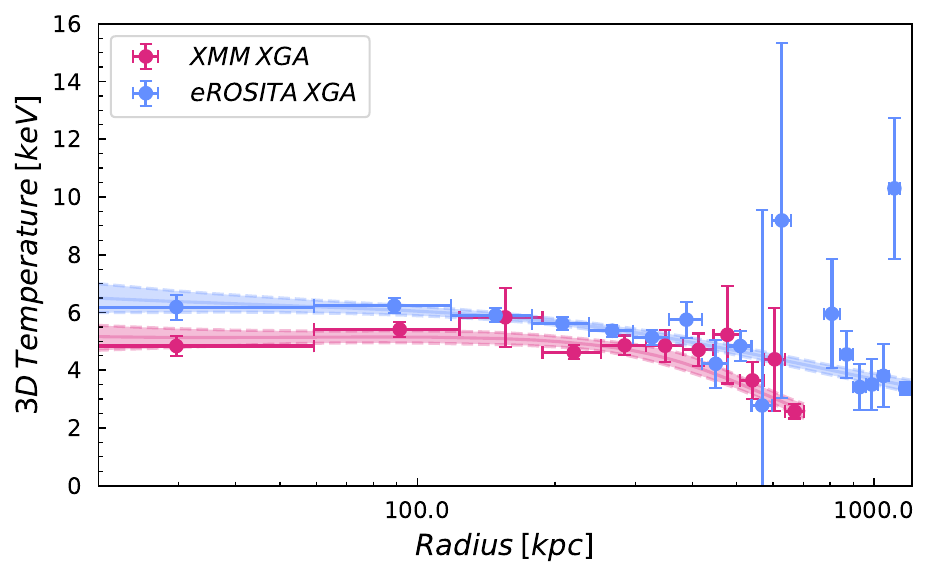}
    \includegraphics[width=0.45\linewidth]{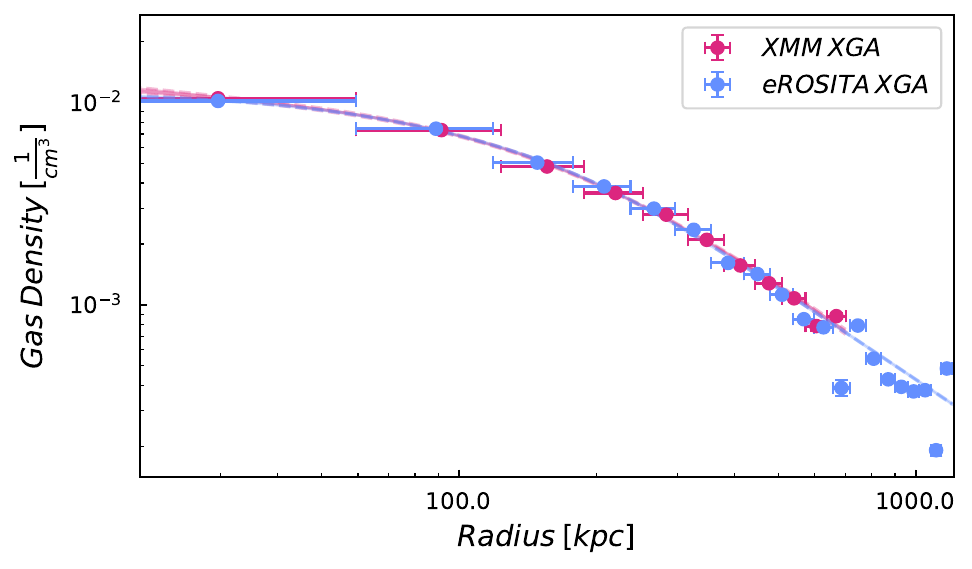}
    \includegraphics[width=0.45\linewidth]{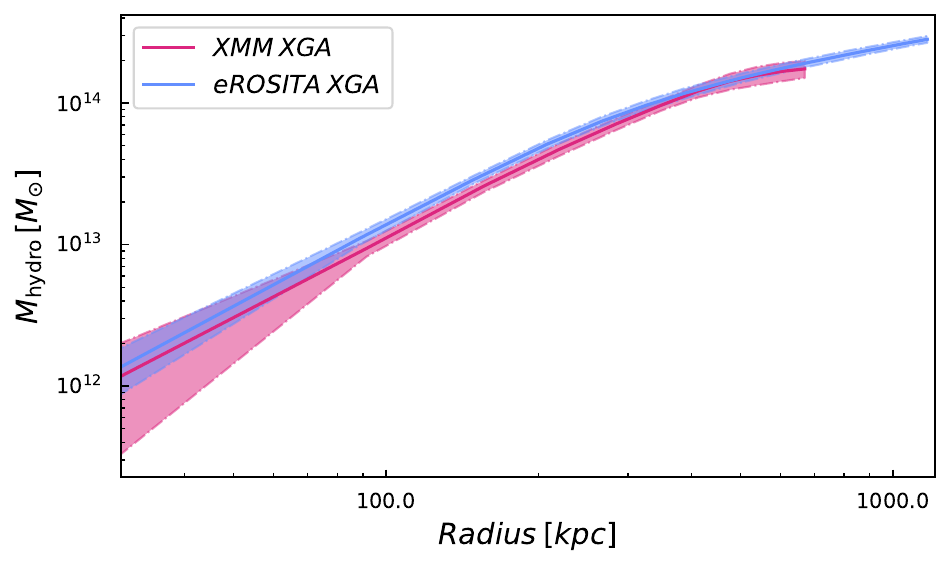}
    \includegraphics[width=0.45\linewidth]{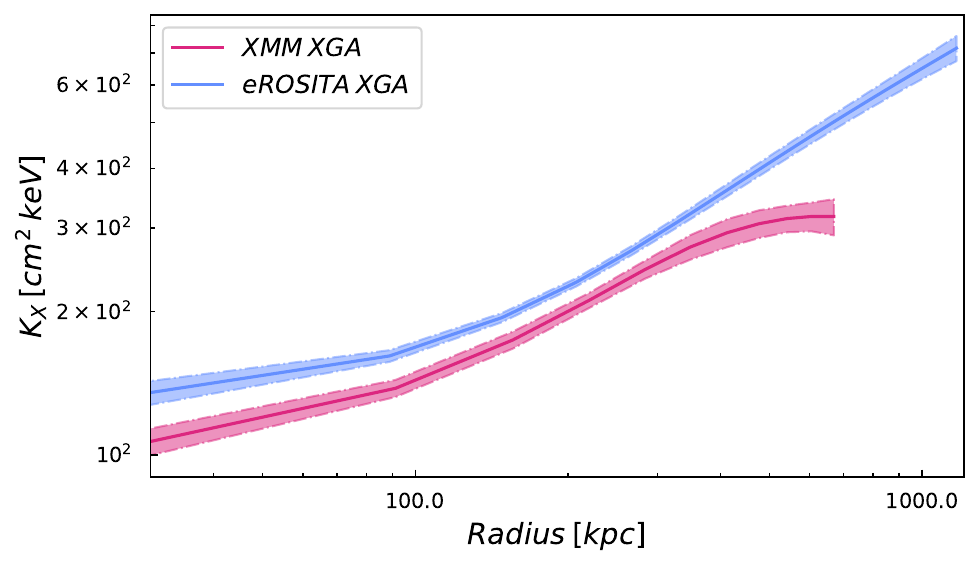}
    \caption{\xmm{} (pink) and \erosita{} (blue) profile measurements of \democluster{}. Upper left: De-projected temperature profile. Upper right: Gas density profile. Estimated from calculations on $N_{\mathrm{APEC}}$, which is obtained from spectral fits. Shaded regions in the upper panels represent model fits to the data. Lower left: Hydrostatic mass profile. Lower right: Entropy profile. Shaded regions in the lower panels represent models derived from the model fits in the  upper panels.}
    \label{fig:profs}
\end{figure*}

We present the entropy profile of \democluster{} in the lower right panel of Figure~\ref{fig:profs}. The general shape of the profile (low in the centre and increasing towards the outskirts) is consistent with previous  measurements of cluster thermodynamic profiles, e.g. \cite{acceptI_entropy, Pratt2010, Ghirardini2019, Riva2024}. A systematic offset between \erosita{} and \xmm{} is evident. This is driven by the corresponding offset in the de-projected temperature profile (upper left of Figure~\ref{fig:profs}), see Equation~\ref{equ:entropy}.


\subsubsection{Summary of \democluster{} Profile Analysis}
 In summary, through comparisons with the \xga{} analysis of the \xmm{} observation of \democluster{}, and with the independent analysis of the same data by \citetalias{Whelan2022}, we conclude that the \xga{} methodology outlined in Section~\ref{subsec:specxga} is reliable when applied to annular spectral analysis of high signal to noise \erosita{} observations made in pointed mode.



\label{subsec:tempprofDis}

\section{Further Validation Tests}
\label{sec:valid}
In this section we validate the methodology outlined in Section~\ref{subsec:specxga} when applied to cluster-wide and core excluded analyses of \erosita{}\footnote{We do not validate \xga{} for \xmm{} analysis, as this was done in {\citetalias{turner2025}}.}
 observations made in survey mode.
For this we have compared \xga{}-derived measurements to values reported in the literature. We use two samples of clusters for validation, as described in Sections~\ref{sec:validation_sample} and~\ref{subsection:des-xcs-ero}, and compare \xga{} measurements of cluster-wide temperatures ($T_{\rm X}$), core excluded temperatures ($T_{\rm X}^{CE}$), 
and soft band (0.5 -2 keV) luminosities ($L_{\rm X, soft}$).

All our comparisons to the literature are fitted with a power law of the form:

\begin{equation}
    \log\left(\frac{Y}{Y_0}\right) = \log(A) + B\log\left(\frac{X}{X_0}\right) \pm \sigma,
\label{equ:polaw}
\end{equation}
where $A$ is the normalisation, $B$ is the slope, $\sigma$ is the scatter, and $X_0$ and $Y_0$ are the normalisation constants for $X$ and $Y$ respectively. All fits to Equation~\ref{equ:polaw} (and Equation~\ref{equ:powlawez} in Section~\ref{sec:scalingrel}) were performed using the {\tt LIRA} \citep[LInear Regression in Astronomy;][]{LIRA2016} package, which is described in detail in \cite{sereno2016}. 


For all \xga{} $T_{\rm X}$, $T_{\rm X}^{CE}$, and $L_{\rm X, soft}$ measurements presented in this section (and in Sections~\ref{sec:XMMCalib} and \ref{sec:scalingrel}, including for $L_{\rm X, bol}$), the following quality criteria are applied for a measurement to be deemed successful:


\begin{itemize}
    \item Measurement errors ($\Delta X$) lie in the range $0< \frac{\Delta X}{X} <0.8$.
    \item 0.5 $<T_{\rm X}<$ 20 keV 
    \item $10^{41} < L_{\rm X} < 10^{46}$~ergs$^{-1}$~cm$^{-2}$
\end{itemize}

The results of the comparisons to the literature are summarised in Table \ref{tab:eroxgacomp}.

\begin{table}
\begin{center}
\caption{Best-fit parameters from the comparison of \xga{} measurements with values from literature. The `Comparison' column lists the literature dataset and property used for comparison. Normalisation and Scatter correspond to $A$ and $\sigma$ respectively in Equation \ref{equ:polaw}, with the slope $B$ fixed at 1, and the values of $X_0$ and $Y_0$ set to 1. The `Sample Size' column indicates the number of clusters used for the respective comparison.}\label{tab:eroxgacomp}
\vspace{1mm}
\begin{tabular}{p{0.3\linewidth}|p{0.2\linewidth}|p{0.2\linewidth}| p{0.1\linewidth}}
\hline
\hline
 Comparison & Normalisation & Scatter & Sample Size \\
\hline
\citetalias{Migkas2024} $T_{\rm X}^{CE}$ & \migkasNORM{} & \migkasSCATTER{} & \migkasSAMP{} \\
\citetalias{Bulbul2024} $T_{\rm X}$ & \bulTNORM{} & \bulTSCATTER{} & \bulTSAMP{} \\
\citetalias{Bulbul2024} $L_{\rm X, soft}$ & \bulLNORM{} & \bulLSCATTER{} & \bulLSAMP{} \\ 
\hline
\hline
\end{tabular}
\end{center}
\end{table}
  
\subsection{Comparison to \citetalias{Migkas2024}}
\label{subsec:M24}
Using \xga{} we have re-analysed the 119 clusters presented in \citetalias{Migkas2024} that had $T_{\rm 
X}^{CE}$ measurements. While the \citetalias{Migkas2024} and \xga{} methodologies are not identical, the approaches have some differences, summarised in Table \ref{tab:m24diff}, and share several similarities which are detailed below:

\begin{itemize}[leftmargin=*, labelindent=2pt, itemsep=1ex]
\item Spectra are extracted from a 0.2-0.5 \rad{} annuli centred on the same positions (provided by \citetalias{Migkas2024}).
\item Analysis uses the combined spectra from all seven telescope modules on board \erosita{}.
\item A \texttt{constant${\times}$tbabs${\times}$apec} model is fit to source spectra.
\item The \cite{aspl} Solar abundance table is used.
\item C-statistics \citep{cash} as the goodness of fit estimator.
\end{itemize}

\begin{table*}
\begin{center}
\caption{Differences in the methodology for measuring $T_{\rm X}^{CE}$ between \xga{} and \citetalias{Migkas2024}.}
\label{tab:m24diff}
\vspace{1mm}
\begin{tabular}{p{0.15\linewidth}|p{0.35\linewidth}|p{0.35\linewidth}}
\hline
\hline
 Method & \citetalias{Migkas2024} & \xga{} \\
\hline
Accounting for the background & Background modelling of a background region of annulus with an inner radii of 1.6 \rad{}, with the outer radii set so that the background area is at least 4 times that of the source region. Fitted simultaneously including components from the Local Hot Bubble, absorbed Milky Way Halo, the cosmic X-ray Background, and the instrumental background & Background subtraction, using an in-field background spectra from a local annulus of 1.05 - 1.5 \rad{} around the analysis region is subtracted from the analysis spectra after the appropriate scaling has taken place\\
\hline
Masking & Custom \erosita{} source detection, which is based upon \cite{erasssrccat} with additional visual inspection & Masks generated by \xapa{} for \xmm{} observations. In the case where the local background region extends beyond the \xmm{} field of view, or no \xmm{} data is available, we default to using masks based on sources detected in \cite{erasssrccat}, described fully in Table \ref{tab:eromasking}\\
\hline
Spectral Fitting Bandwidth & 0.7 - 7 keV & 0.3 - 7.9 keV\\
\hline 
Metallicity Parameter & Frozen to the value obtained in \cite{migkas2020} & Frozen to 0.3Z$\odot$ \\
\hline
$n_H$ & Frozen to the value obtained in \cite{migkas2020} & Frozen to the value obtained from the HEASoft \texttt{nh} command, taken from \cite{nh} \\
\hline
\textsc{XSPEC} version & 12.12.0 & 12.14.1 \\
\hline
\hline
\end{tabular}
\end{center}
\end{table*}

\begin{figure}
    \centering
    \includegraphics[width=1.\columnwidth]{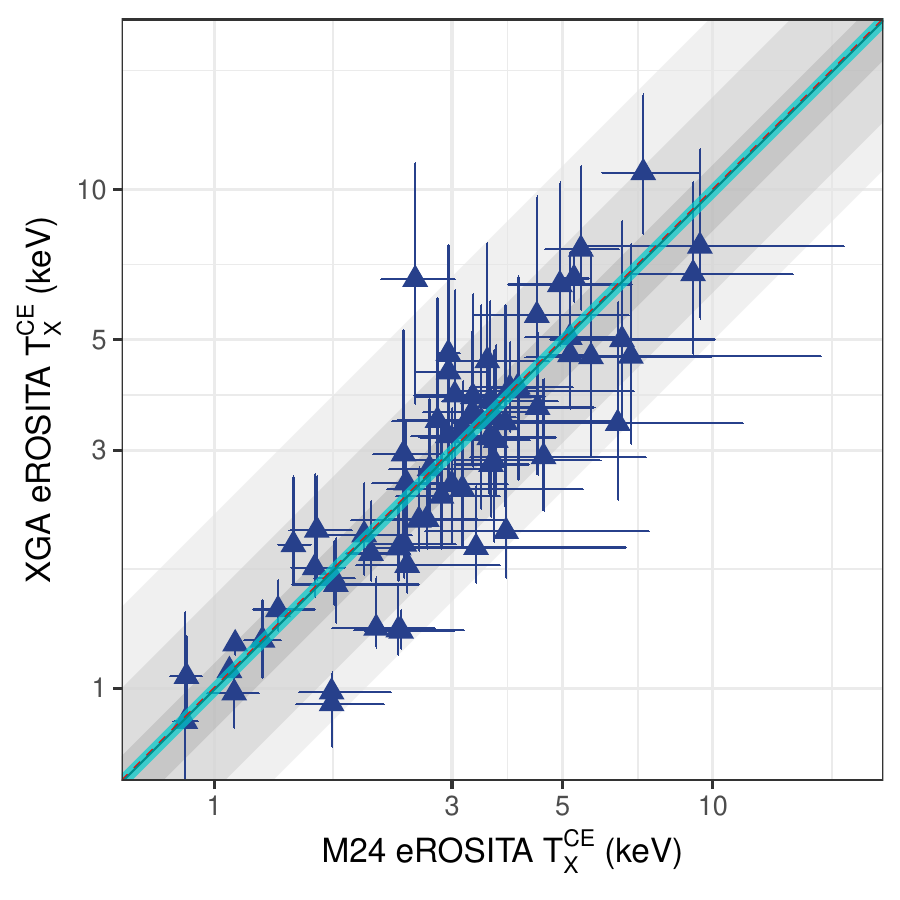}
    \caption{The \xga{} measured $T_{\rm X}^{CE}$ versus those measured by \citetalias{Migkas2024} using the \erosita{} telescope. The best-fit to the relation is given by the blue line, the cyan band around it represents the 1$\sigma$ error on the line, and the red dashed line represents the 1:1 relation. The grey shaded regions from darkest to lightest represent the 1$\sigma$ 2$\sigma$ and 3$\sigma$ intrinsic scatter.} 
    \label{fig:m24erotemp}
\end{figure}

As outlined in Section~\ref{sec:validation_sample}, \citetalias{Migkas2024} measures T$_{\rm X}^{CE}$ for 119 galaxy clusters with \erosita{}. Using the \xga{} methodology as described in Section~\ref{subsec:specxga} and the quality cuts detailed at the beginning of this Section, we estimate $T_{\rm X}^{CE}$ for \migkasSAMP{} of the 119 clusters from \citetalias{Migkas2024}. Figure~\ref{fig:m24erotemp} shows both studies are in close agreement, and a fit to the data (assuming Equation~\ref{equ:polaw} with  fixed slope of $B=1$) yields $A$=\migkasNORM{} (represented by the cyan line and shaded region in Figure~\ref{fig:m24erotemp}), which is consistent with a 1:1 relation. When the slope is left as a free parameter, the best fit value is \migkasSLOPE{}.


With regard to the 51 (of 119) \citetalias{Migkas2024} clusters not included in the fit, most have \xga{} errors larger than our $\bigr( \frac{\Delta T}{T}\bigl)<0.8$ limit: 41 have errors in the $0.8<\bigr( \frac{\Delta T}{T}\bigl)<1.5$ range. The errors quoted by \citetalias{Migkas2024} on their 119 $T_{\rm X}^{CE}$ measurements range from 0.02 to 1.51. The remaining 10 (of 51) either did not return a value within the \xga{} allowed range of $0.5 <T_{\rm X}< 20$ keV or had a error greater than  $\bigr( \frac{\Delta T}{T}\bigl)=1.5$.


\subsection{Comparison to \citetalias{Bulbul2024}}
\label{subssec:primcat}
Using \xga{} we perform spectral fits up to \rad{} to measure $T_{\rm X}$ and $L_{\rm X, soft}$ for all 114 clusters in the DES-XCS-eRO sample  (fully described in Section~\ref{subsection:des-xcs-ero}).

The methods for X-ray analysis in \xga{} and \citetalias{Bulbul2024} differ significantly. In \citetalias{Bulbul2024}, measurements are performed using the MultiBand Projector in 2D \citep[{\tt MBProj2D},][]{mbproj2d2018} code. {\tt MBProj2D} simultaneously fits several X-ray images across multiple energy bands from the data to forward model generated X-ray images to obtain various parameter measurements and profiles, such as $T_{\rm X}$ and hydrostatic mass (among others). For the analysis in \citetalias{Bulbul2024}, images and exposure maps of size 8 \rad{} x 8 \rad{} in energy bands of [0.3-0.6], [0.6-1.0], [1.0-1.6], [1.6-2.2], [2.2-3.5], [3.5-5.0] and [5.0-7.0]~keV, are used as input into the {\tt MBProj2D} code. They mask out faint contaminating sources, and model the brighter ones. To account for the background, it is modelled using a local region. For a full description of the {\tt MBProj2D} methodology, see Appendix A of \citetalias{Bulbul2024}, and its application for the \citetalias{Bulbul2024} cluster parameter measurements is detailed in Section 5.1 of \citetalias{Bulbul2024}.  The \xga{} analysis in this study and the analysis performed in \citetalias{Bulbul2024} have the following analysis choices in common; fixing the metal abundance at $Z=0.3Z_\odot$; use the \cite{aspl} Solar abundance table; use \texttt{nh} values from \cite{nh}; and use C-stat for the goodness of fit estimator.

\begin{figure}
    \centering
    \includegraphics[width=1.\columnwidth]{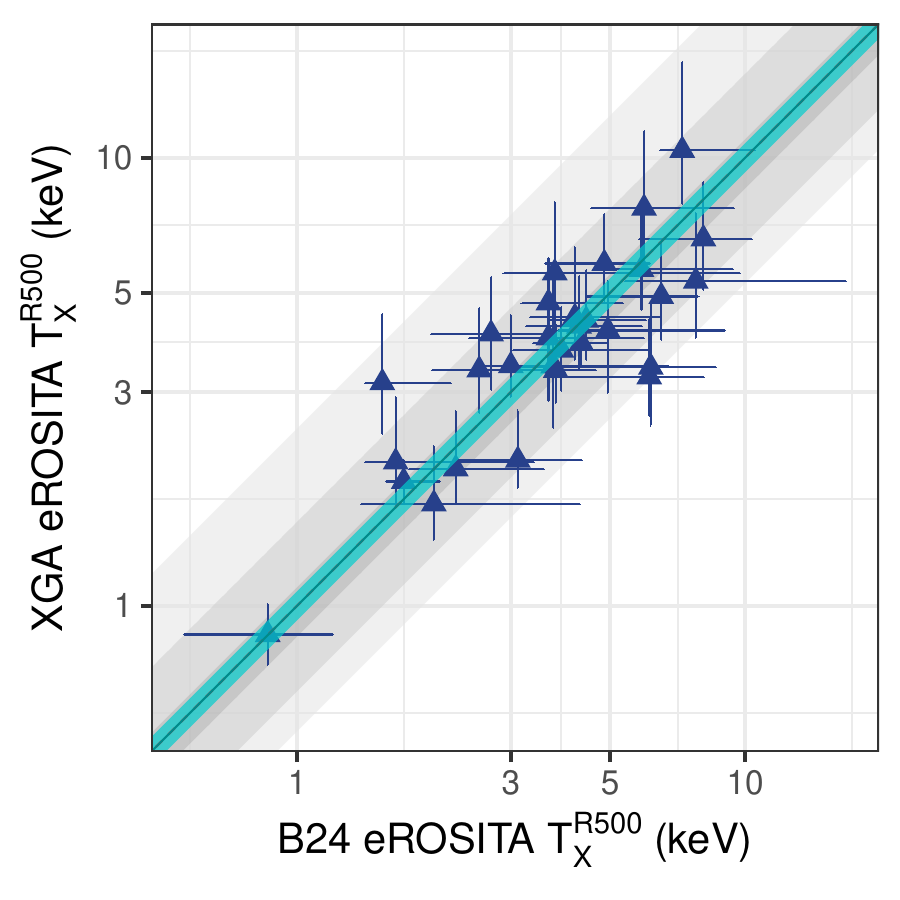}
    \caption{The \xga{} measured $T_{\rm X}$ versus those measured by \citetalias{Bulbul2024} for the \erosita{} telescope. The blue line is the best fit line, the cyan band around it represents the 1$\sigma$ error on the line, and the red dashed line represents the 1:1 relation. The grey shaded regions from darkest to lightest represent the 1$\sigma$ 2$\sigma$ and 3$\sigma$ scatter. }
    \label{fig:B24comp}
\end{figure}

Applying the methodology described in Section \ref{subsec:specxga}, \xga{} measures $T_{\rm X}$ for 47 of the 114 DES–XCS–eRO clusters using \erosita{} data. \citetalias{Bulbul2024} provides $T_{\rm X}$ values for 50 of the DES-XCS-eRO sample. Of the 47 clusters measured by \xga{}, and 50 measured by \citetalias{Bulbul2024}, there are \bulTSAMP{} clusters in common with successful\footnote{To be ``successful'', the \xga{} fits had to meet these criteria: $ \frac{\Delta T_{\rm X}}{T_{\rm X}} <0.8$, and 0.5 $<T_{\rm X}<$ 20 keV, see Section~\ref{sec:valid}).} $T_{\rm X}$ measurements. 
The comparison of the \bulTSAMP{} clusters in common is shown in Figure~\ref{fig:B24comp} (blue points) and a fit (assuming Equation~\ref{equ:polaw} with  fixed slope of $B=1$) to the data is given by the blue line and cyan shaded region highlighting the 1$\sigma$ uncertainty. We find $A$=\bulTNORM{}, consistent with a 1:1 relation. When the slope is left as a free parameter, the best fit value is \bulTSLOPE{}.

\begin{figure}
    \centering
    \includegraphics[width=1.\columnwidth]{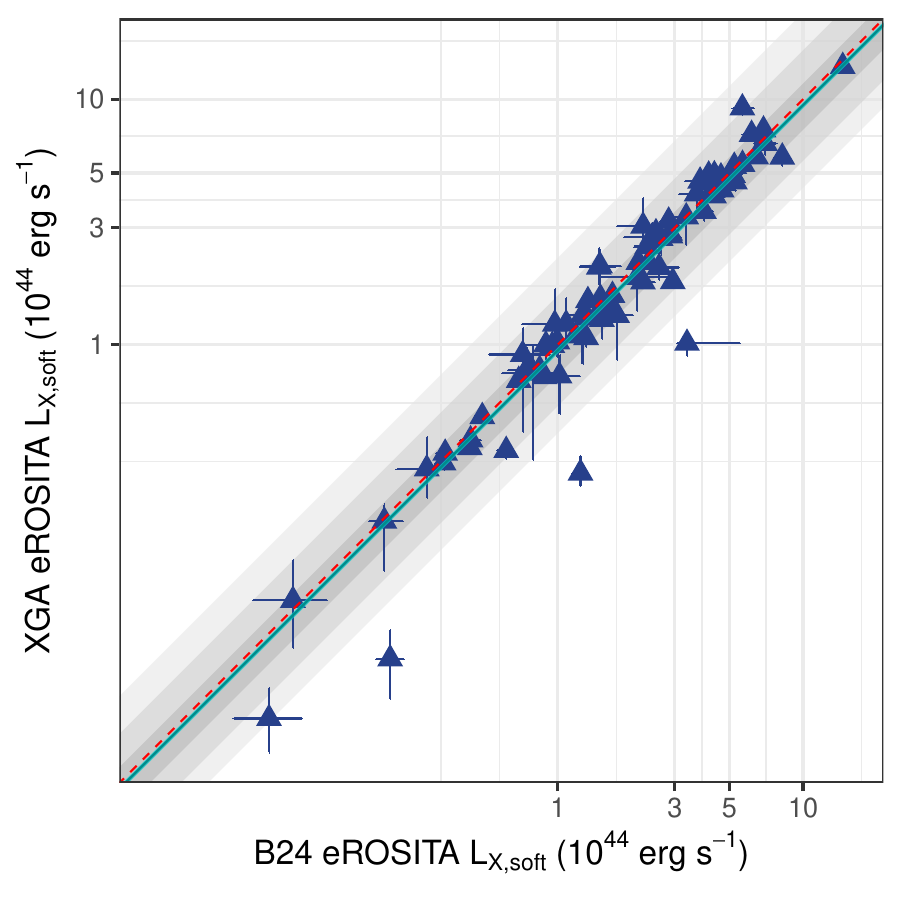}
    \caption{The \xga{} measured $L_{\rm X, soft}$ in the 0.5 - 2 keV band versus those measured by \citetalias{Bulbul2024}. The blue line is the best fit line, the cyan band around it represents the 1$\sigma$ error on the line (note that the cyan band is difficult to distinguish in the plot due to the tight constraints in the fit), and the red dashed line represents the 1:1 relation. The grey shaded regions from darkest to lightest represent the 1$\sigma$ 2$\sigma$ and 3$\sigma$ scatter.}
    \label{fig:B24erolumcomp}
\end{figure}

We also constrain $L_{\rm X, soft}$ in the 0.5 - 2 keV band. \citetalias{Bulbul2024} measures 109 $L_{\rm X, soft}$ values of the 114 DES-XCS-eRO clusters.  From the \xga{} analysis, we measure 76 $L_{\rm X, soft}$ values for the DES-XCS-eRO sample (all in common with \citetalias{Bulbul2024}).  
We present the comparison of these measurements in Figure~\ref{fig:B24erolumcomp} (blue points). We fit Equation \ref{equ:polaw} to the data with afixed slope of $B=1$. We find $A$=\bulLNORM{}, consistent with a 1:1 relation. When the slope is left as a free parameter, the best fit value is \bulLSLOPE{}.



\subsubsection{Summary of Sample Comparison Analysis}
\label{subsec:validDis}

In summary, we have validated that the methodology outlined in Section~\ref{subsec:specxga} is reliable when applied to the analysis of \erosita{} observations of clusters made in survey mode, specifically the measurement of luminosities and of temperatures (both global (<\rad{}) and core excised).
We have done this by comparison to \xga{}  measurements published values in \citetalias{Migkas2024} and \citetalias{Bulbul2024}.

\section{Cross-Calibration of \xmm{} and \erosita{}}
\label{sec:XMMCalib}

We have shown above that the \xga{} processing of \erosita{} observations is robust (Sections~\ref{subsec:profxga} and \ref{sec:valid}). Similarly the \xga{} processing of \xmm{} observations was validated in \citetalias{turner2025}. Therefore, we can now confidently compare the \xga{} analysis of clusters observed by both telescopes to establish their relative cross calibration. For this, we make measurements of $T_{\rm X}$, $L_{\rm X, bol}$, and $L_{\rm X, soft}$ for the DES-XCS-eRO sample (outlined in Section~\ref{subsection:des-xcs-ero}) using \erosita{} and \xmm{}.
For all comparisons between the two telescopes we fit a power law described by Equation \ref{equ:polaw}. We perform fits both with the slope fixed to 1 and with the slope free to vary. The latter allows us to investigate the dependence of any measured offsets with the X-ray property in question. We present the fit parameters in Table \ref{tab:tel_comp}.

\begin{table*}
\begin{center}
\caption{Fit values comparing \erosita{} and \xmm{} measurements of $L_{\rm X, soft}$, $L_{\rm X, bol}$, and $T_{\rm X}$. The `Normalisation', `Slope', and `Scatter' columns are the value of $A$, $B$, and $\sigma$ respectively in Equation \ref{equ:polaw}, with the `$X_0$' and `$Y_0$' columns defined in the same equation. $T_{\rm X}$ results are for the sample with a minimum count threshold of 150, as measured by \citetalias{Bulbul2024}. The $L_{\rm X, bol}$ and $L_{\rm X, soft}$ relations with a sample size of 37 use an identical set of clusters to the $T_{\rm X}$ relation.}\label{tab:tel_comp}
\vspace{1mm}
\begin{tabular}{p{0.12\linewidth}|p{0.12\linewidth}|p{0.12\linewidth}| p{0.12\linewidth}|p{0.08\linewidth}| p{0.08\linewidth}|p{0.08\linewidth}| p{0.08\linewidth}}
\hline
\hline
 Comparison & Normalisation & Slope & Scatter & Sample Size & $X_0$ & $Y_0$ & Figure~\\
\hline
\multicolumn{8}{c}{\textit{FIXED SLOPE}} \\
[0.5em]
$L_{\rm X, soft}$ & \LsoftNORMfix{} & 1$\pm$0 & \LsoftSCATTERfix{} & \LsoftSAMP{} & 1.0 & 1.0 & Left panel of Figure~\ref{fig:lx_cross_calib} \\
$L_{\rm X, soft}$ & \LsoftNORMfixT{} & 1$\pm$0 & \LsoftSCATTERfixT{} & \SAMPT{} & 1.0 & 1.0 & - \\
$L_{\rm X, bol}$  & \LbolNORMfix{} & 1$\pm$0 & \LbolSCATTERfix{} & \LbolSAMP{} & 1.0 & 1.0 & Right panel of Figure~\ref{fig:lx_cross_calib} \\
$L_{\rm X, bol}$  & \LbolNORMfixT{} & 1$\pm$0 & \LbolSCATTERfixT{} & \SAMPT{} & 1.0 & 1.0 & - \\
$T_{\rm X}$ & \TNORMfix{} & 1$\pm$0 & \TSCATTERfix{} & \SAMPT{} & 1.0 & 1.0 & Figure~\ref{fig:b24TempsXgaEroXgaXmmctscut} \\
[0.5em]
\multicolumn{8}{c}{\textit{FREE SLOPE}} \\
[0.5em]
$L_{\rm X, soft}$  & \LsoftNORM{} & \LsoftSLOPE{} & \LsoftSCATTER{} & \LsoftSAMP{} & 2.1 & 2.6 & Left panel of Figure~\ref{fig:lx_cross_calib} \\
$L_{\rm X, soft}$ & \LsoftNORMT{} & \LsoftSLOPET{} & \LsoftSCATTERT{} & \SAMPT{} & 2.1 & 2.6 & - \\
$L_{\rm X, bol}$ & \LbolNORM{} & \LbolSLOPE{} & \LbolSCATTER{} & \LbolSAMP{} & 4.4 & 7.8 & Right panel of Figure~\ref{fig:lx_cross_calib} \\
$L_{\rm X, bol}$ & \LbolNORMT{} & \LbolSLOPET{} & \LbolSCATTERT{} & \SAMPT{} & 4.4 & 7.8 & - \\
$T_{\rm X}$ & \TNORM{} & \TSLOPE{} & \TSCATTER{} & \SAMPT{} & 4.2 & 5.0 & Figure~\ref{fig:b24TempsXgaEroXgaXmmctscut} \\
\hline
\hline
\end{tabular}
\end{center}
\end{table*}

\subsection{Cross-calibration of temperature}
\label{subsec:tx_crosscal}

\begin{figure}
    \centering
    \includegraphics[width=1.\columnwidth]{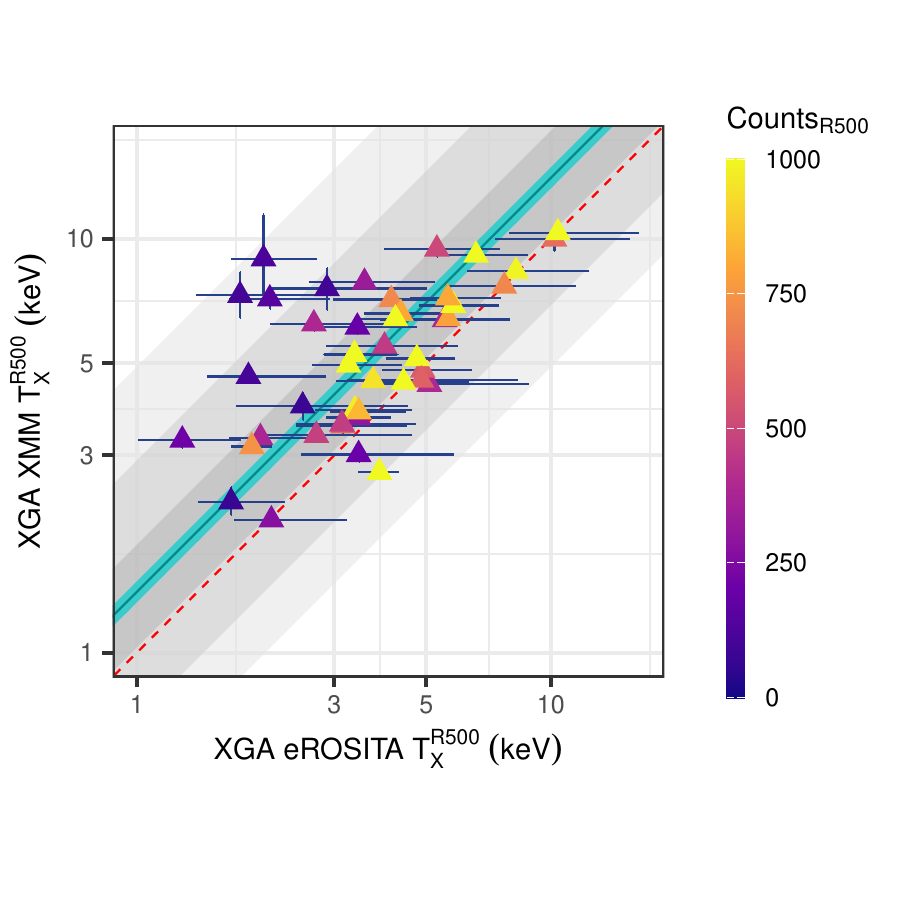}
    \caption{The \xga{} \erosita{} measured $T_{\rm X}$ values for versus \xga{} \xmm{} measured. The blue line is the best fit line, the cyan band around it represents the 1$\sigma$ error on the line, and the red dashed line represents the 1:1 relation. The grey shaded regions from darkest to lightest represent the 1$\sigma$ 2$\sigma$ and 3$\sigma$ scatter. The points are coloured according to their \erosita{} counts, as reported in the ``CTS500'' column in \citetalias{Bulbul2024}. We limit the range of the colour bar from 0 - 1000 counts.}
    \label{fig:b24TempsXgaEroXgaXmm}
\end{figure}

\xga{} successfully$^{20}$ measures global ($<$\rad) $T_{\rm X}$ values for 47 and 103 clusters with \erosita{} and \xmm{} respectively. 
There are 44 clusters with successful measurements by both telescopes.  We find that \xmm{} typically measures higher $T_{\rm X}$ values than \erosita{}, finding a normalisation of $A$=1.40$\pm$0.08, with the slope fixed at 1, see Figure~\ref{fig:b24TempsXgaEroXgaXmm}. When 
allowing the slope to vary, we find an offset of 40$\pm$10\% for clusters measured by \erosita{} at 3~keV, and an offset of 20$\pm$20\% for clusters measured by \erosita{} at 10 keV.  

Previous calibrations of galaxy cluster observations with \erosita{} and \xmm{} have also reported that \xmm{} measures higher $T_{\rm X}$  values than \erosita{}, notably by \citetalias{Turner2022} and \citetalias{Migkas2024}.  \citetalias{Turner2022} found \xmm{} measured $T_{\rm X}$ values 25$\pm$9\% higher than \erosita{}. \citetalias{Migkas2024} finds \xmm{} measures $T_{\rm X}$ 14\% and 32\% higher for clusters measured at 3~keV and 10~keV with \xmm{} respectively.  While our results are not in significant tension with these previous studies (due to the large errors on the measured offset), the difference warrants further scrutiny.  In Figure~\ref{fig:b24TempsXgaEroXgaXmm}, the symbols are colour-coded by \erosita{} counts, which are taken directly from \citetalias{Bulbul2024} in the ``CTS500'' column. It can be seen that clusters with low counts ($\lesssim$ 250) in \erosita{} tend to lie the furthest from the 1:1 relation. Notably, at an \xmm{} estimated $T_{\rm X}\sim$7~keV, the \erosita{} measured values range from $\sim$2 to 10~keV. This large range in \erosita{} measured $T_{\rm X}$ values is likely artificial, due to the lower signal to noise of \erass{} detections compared to \xmm{}: for all 47 clusters, the \xmm{} source counts are higher, with a median factor of 52, than those from \erosita{}.

\begin{figure}
    \centering
    \includegraphics[width=1.\columnwidth]{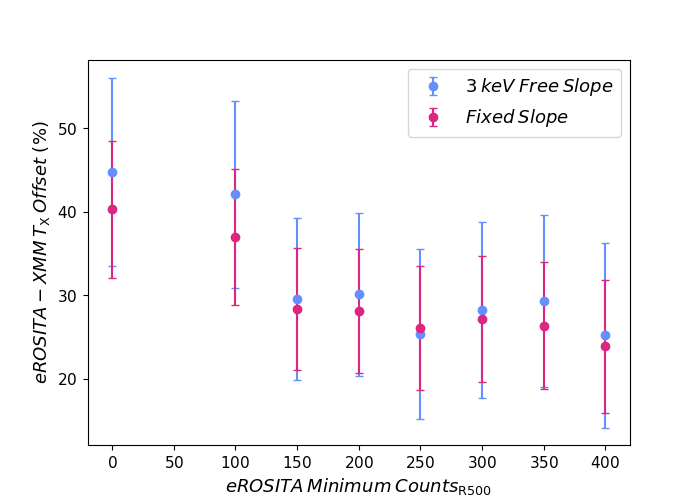}
    \caption{Trend in the \erosita{}–\xmm{} $T_{\rm X}$ offset as a function of the lower count threshold applied to the cluster sample. The offset is defined as the difference between the \xmm{} and \erosita{} $T_{\rm X}$ measurement, divided by the \erosita{} $T_{\rm X}$ measurement, given as a percentage. The offset fitted with a fixed slope is shown in pink, while the offset for a cluster with $T_{\rm X} = 3$keV measured by \erosita{} with a varying slope is shown in blue.}
    \label{fig:b24offsets}
\end{figure}

In order to quantify the effect of \erosita{} counts on the \xmm{} offset, we excluded clusters under a minimum \erosita{} count threshold and repeated the $T_{\rm X}$ comparison. We repeated the fits for 8 minimum \erosita{} count thresholds, starting at 0 counts (i.e., including all clusters in the DES-XCS-eRO sample), and up to a minimum \erosita{} measured count of 400. The \erosita{} - \xmm{} $T_{\rm X}$ offset as a function of count threshold is shown in Figure~\ref{fig:b24offsets}. It is found that for a minimum \erosita{} count of $>$150, the measured offsets fluctuate around an offset of \textasciitilde30\%. Below a minimum threshold of 150 \erosita{} counts, the value of the offset increases. We visually inspected the spectra of clusters with $<$150 counts, which revealed that there was negligible signal beyond 2 keV, where the sensitivity of \erosita{} drops off steeply. This lack of detected higher energy photons causes \erosita{} to erroneously measure $T_{\rm X}$ $\lesssim$ 3 keV. We therefore draw the conclusion that measurements of $T_{\rm X}$ with \erosita{} using the \xga{} method are only reliable for spectra generated within regions that have counts of 150 and above. 

\begin{figure}
    \centering
    \includegraphics[width=1.\columnwidth]{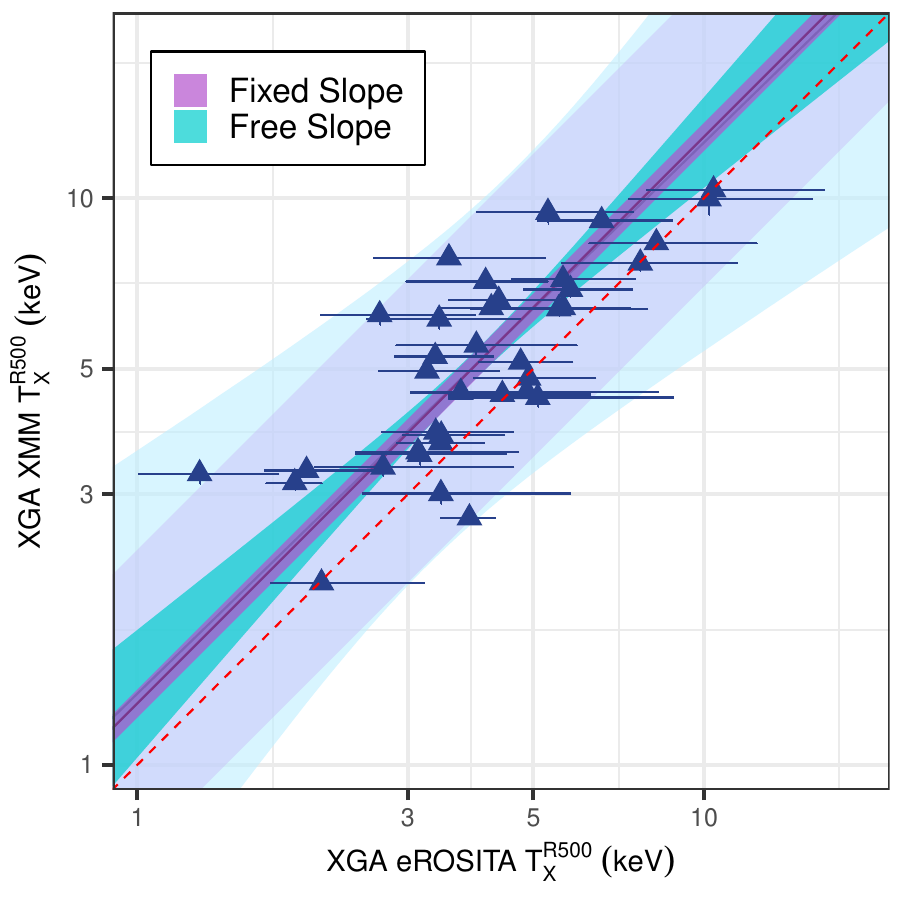}
    \caption{The \xga{} \erosita{} measured $T_{\rm X}$ values versus \xga{} \xmm{} measured ones, cutting the sample by a minimum of 150 counts measured by \erosita. For the fit with a varying slope: the blue line is the best fit line, the cyan band around it represents the 1$\sigma$ error on the line, and the light blue shaded region represents the 1$\sigma$ scatter. For the fit with a fixed slope: the dark purple line is the best fit line, the purple band around it represents the 1$\sigma$ error on the line, and the light purple shaded region represents the 1$\sigma$ scatter. The red dashed line represents the 1:1 relation.}
    \label{fig:b24TempsXgaEroXgaXmmctscut}
\end{figure}

Figure~\ref{fig:b24TempsXgaEroXgaXmmctscut} shows the \erosita{} - \xmm{} $T_{\rm X}$ offset for clusters with more than 150 counts in \erosita{}. We find that \xmm{} measures higher $T_{\rm X}$ values than \erosita{}, and recover $A=$~\TNORMfix{} for a fixed slope. We also allow the slope to vary in the fit, measuring $B = $ \TSLOPE{}. Therefore, unlike \citetalias{Migkas2024}, we find no evidence of a $T_{\rm X}$-dependent offset. We attribute the discrepancy with \citetalias{Migkas2024} to the different samples used in the two works: 58\% of the \citetalias{Migkas2024} sample have an \erosita{} $T_{\rm X} > 3$ keV, compared to 87\% in this work. The slope measured here is therefore driven largely by hotter clusters, whereas the slope in \citetalias{Migkas2024} is weighted more evenly between the cool and hot cluster populations. Indeed, our measured offset of 28$\pm$7\% is consistent with the 32\% offset measured at 10 keV in \citetalias{Migkas2024}.



\subsection{Cross-calibration of luminosity}
\label{subsec:lx_crosscal}

\begin{figure*}
    \centering
    \includegraphics[width=0.45\linewidth]{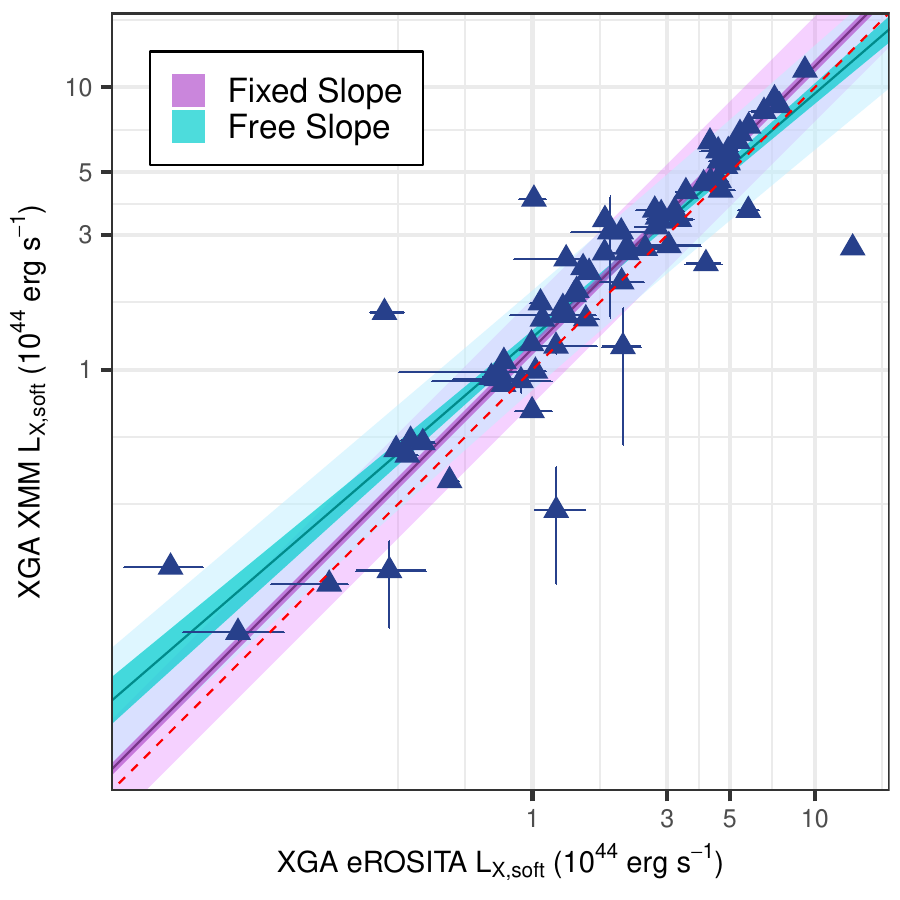}
    \includegraphics[width=0.45\linewidth]{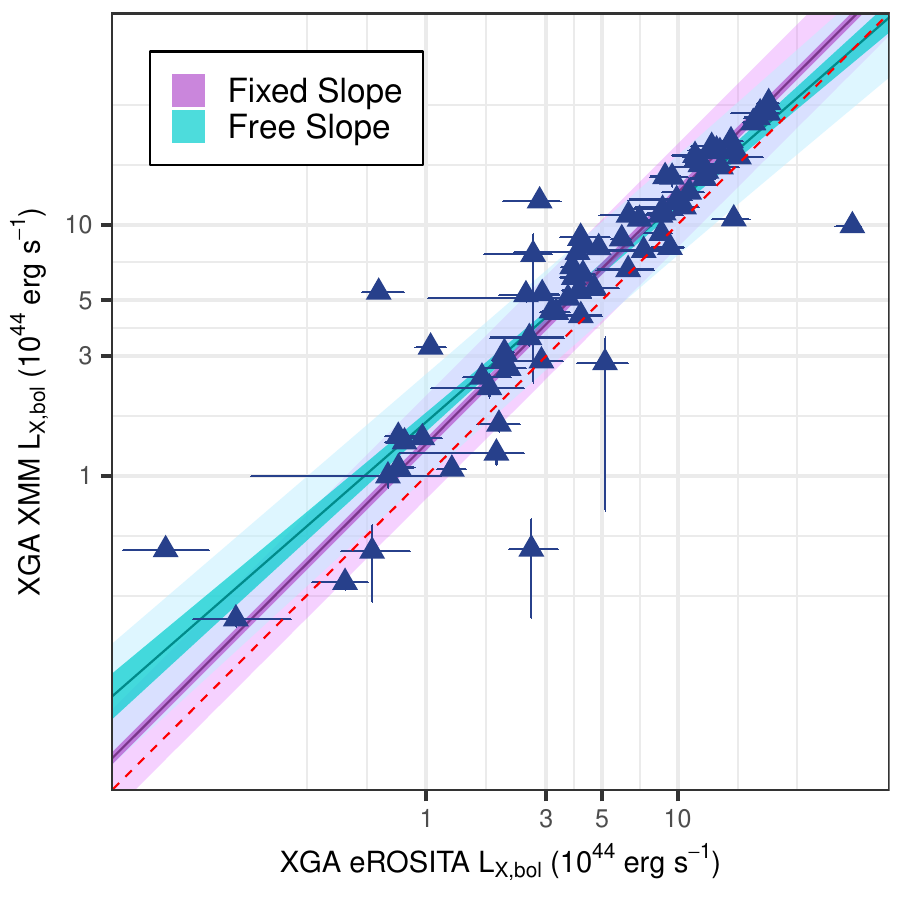}
    \caption{Comparison of the \LsoftSAMP{} and \LbolSAMP{} \xga{} measured $L_{\rm X, soft}$ and $L_{\rm X, bol}$ values for the DES-XCS-eRO sample using \erosita{} and \xmm{} for the left and right panels respectively. For the fit with a varying slope: the blue line is the best fit line, the cyan band around it represents the 1$\sigma$ error on the line, and the light blue shaded region represents the 1$\sigma$ scatter. For the fit with a fixed slope: the dark purple line is the best fit line, the purple band around it represents the 1$\sigma$ error on the line, and the light purple shaded region represents the 1$\sigma$ scatter. The red dashed line represents the 1:1 relation.}
    \label{fig:lx_cross_calib}
\end{figure*}

\xga{} measures $L_{\rm X, soft}$ for 76 and 106 clusters with \erosita{} and \xmm{} respectively, providing \LsoftSAMP{} clusters with measurements from both observatories. For the bolometric bandwidth (0.1-10.0 keV), one additional \xmm{} measurement was obtained. Figure~\ref{fig:lx_cross_calib} shows the \erosita{} measurements of $L_{\rm X, soft}$ and $L_{\rm X, bol}$ compared to \xmm{}. In both cases we find that \xmm{} measures higher luminosity values than \erosita{}, finding a normalisation of $A$=\LsoftNORMfix{} and $A$=\LbolNORMfix{} for $L_{\rm X, soft}$ and $L_{\rm X, bol}$ respectively.

To test for an luminosity dependence in the offset, we allow the slope parameter ($B$) of Equation~\ref{equ:polaw} to vary. For $L_{\rm X, soft}$, we obtain $B$ = \LsoftSLOPE{}, corresponding to an \xmm{} $L_{\rm X, soft}$ that is $18 \pm 5$ \% higher than \erosita{} at $L_{\rm X,soft} = 2\times10^{44}\ \mathrm{erg\ s^{-1}}$, and consistent with parity at $L_{\rm X,soft} = 8\times10^{44}\ \mathrm{erg\ s^{-1}}$. For $L_{\rm X, bol}$, allowing the slope to vary yields $B$ = \LbolSLOPE{}, implying \xmm{} $L_{\rm X, bol}$ values that are $50 \pm 10$\% higher than \erosita{} at $2\times10^{44}\ \mathrm{erg\ s^{-1}}$, and $26 \pm 8$\% higher at $8\times10^{44}\ \mathrm{erg\ s^{-1}}$.  

Bolometric X-ray luminosities scale as $L_{\rm X, bol} \propto T_{\rm X}^2 $, while soft-band luminosities follow a weaker dependence, $L_{\rm X, soft} \propto T_{\rm X}^{3/2}$ \citep[see Section 2.1 in][for details]{Lovisari2021}. This amplifies any systematic offset in $T_{\rm X}$ between \erosita{} and \xmm{}, and could naturally contribute to the larger offset observed in $L_{\rm X, bol}$ relative to $L_{\rm X, soft}$. A further contributing factor may arise from the effects of modelling a multi-temperature ICM with a single-temperature model, the impact of which is discussed in detail in Section~\ref{sec:disc:multiICM}.

To explore the dependence of the luminosity offset on the $T_{\rm X}$ constraint, we repeated the luminosity comparisons for the subsample of \SAMPT{} clusters which had a successfully$^{20}$  measured $T_{\rm X}$ value (Section~\ref{subsec:tx_crosscal}). We find no significant difference in the fit parameters between the well constrained $T_{\rm X}$ subsample and the full luminosity subsamples, as shown in Table \ref{tab:tel_comp}, no fit values vary by more than 2$\sigma$.

\citetalias{Turner2022} compared $L_{\rm X, soft}$ for \erosita{} eFEDS clusters and \xmm{}, finding them to be consistent. It is worth noting that the \citetalias{Turner2022} comparison was made within a fixed 500~kpc aperture, whereas the comparison presented in this work uses $R_{500}$ taken from \citetalias{Bulbul2024}. A further factor that may contribute to any discrepancy is the depth of the eFEDS data, which is a factor 8$\times$ deeper than \erass{}. In addition, \citetalias{Bulbul2024} compared their \erosita{} \erass{} $L_{\rm X, soft}$ measurements to those from \chandra{}, and found the \erosita{} values to be systematically lower, by approximately 15\%. This offset is consistent with the $19\pm6$\% difference measured in this work.

\section{X-ray - Optical Scaling Relations}
\label{sec:scalingrel}

Having established $L_{\rm X}$  and $T_{\rm X}$ offsets between \xmm{} and \erosita{} (Section~\ref{sec:XMMCalib}), we can now explore whether it will be possible to combine \xmm{} and \erosita{} observations to improve cluster cosmology studies, specifically through a better understanding of the X-ray - richness relations $L_{\rm X, soft}$-$\lambda_{RM}$ and $T_{\rm X}$-$\lambda_{\rm RM}$. The \xmm{} and \erosita{} relations are derived using identical subsets of DES-XCS-eRO clusters, namely, those for which successful$^{20}$  measurements were obtained from both telescopes.

For this, we modify Equation \ref{equ:polaw} by accounting for redshift evolution, $E(z)$, giving the following:
\begin{equation}
    \mathrm{log}\biggr(E(z)^{\gamma} \cdot \frac{Y}{Y_0}\biggl) \ = \ \mathrm{log}(A) + B \cdot \mathrm{log}\biggr(\frac{\lambda _{RM}}{\lambda _{0}}\biggl) \ \pm \ \sigma,
\label{equ:powlawez}
\end{equation}
where $E(z) = \sqrt{\Omega_M(1 + z)^3 + \Omega _\Lambda}$, assuming $\Omega _M$ = 0.3 and $\Omega _\Lambda$ = 0.7. We set $\gamma$ to $-1$ and $-\frac{2}{3}$ for the $L_{\rm X, soft}$-$\lambda_{RM}$ and $T_{\rm X}$-$\lambda_{\rm RM}$ relations respectively, which come from the self-similar expectation. We set $Y_0$ and $\lambda_{0}$ to the median values of the data, which are listed in Table \ref{tab:lam-both}, alongside the fit results.

\begin{table*}
\begin{center}
\caption{Fit values of the $L_{\rm X, soft}$-$\lambda_{RM}$ and $T_{\rm X}$-$\lambda_{\rm RM}$  relations, measured with \xga{} using \erosita{} and \xmm{} for the same cluster sample. The `Normalisation', `Slope', and `Scatter' column are the values of $A$, $B$, and $\sigma$ respectively in Equation \ref{equ:powlawez}. For the $L_{\rm X, soft}$-$\lambda_{RM}$ relation, $Y_0$ and $Y$ correspond to $L_{X0}$ and $L_{\rm X, soft}$, while for the $T_{\rm X}$-$\lambda_{\rm RM}$ relation, they correspond to $T_{X0}$ and $T_{\rm X}$. The fitted relations are shown in Figure~\ref{fig:richness_scalrel}. The $L_{\rm X, soft}$ relations with a sample size of 37 use an identical set of clusters to the $T_{\rm X}$ relation.}\label{tab:lam-both}
\vspace{1mm}
\begin{tabular}{p{0.1\linewidth} | p{0.1\linewidth}|p{0.15\linewidth}|p{0.15\linewidth}|p{0.15\linewidth}|p{0.1\linewidth}}
\hline
\hline
 Relation & Telescope & Normalisation & Slope & Scatter & Sample Size \\
\hline
\multicolumn{6}{c}{$\lambda_{0} = 80, L_{X0} = 2.1$} \\
[0.5em]
$L_{\rm X, soft}$-$\lambda_{RM}$ & \erosita{} & \eRICHLsoftNORM & \eRICHLsoftSLOPE & \eRICHLsoftSCATTER & \RICHLSAMP \\
$L_{\rm X, soft}$-$\lambda_{RM}$ & \erosita{} & \eRICHLsoftNORMT & \eRICHLsoftSLOPET & \eRICHLsoftSCATTERT & \RICHTSAMP \\
$L_{\rm X, soft}$-$\lambda_{RM}$ & \xmm{} & \xRICHLsoftNORM & \xRICHLsoftSLOPE & \xRICHLsoftSCATTER & \RICHLSAMP \\
$L_{\rm X, soft}$-$\lambda_{RM}$ & \xmm{} & \xRICHLsoftNORMT & \xRICHLsoftSLOPET & \xRICHLsoftSCATTERT & \RICHTSAMP \\
[0.5em]
\multicolumn{6}{c}{$\lambda_{0} = 96, T_{X0} = 3.9$} \\
[0.5em]
$T_{\rm X}$-$\lambda_{\rm RM}$ & \erosita{} & \eRICHTNORM & \eRICHTSLOPE & \eRICHTSCATTER & \RICHTSAMP \\
$T_{\rm X}$-$\lambda_{\rm RM}$ & \xmm{} & \xRICHTNORM & \xRICHTSLOPE & \xRICHTSCATTER & \RICHTSAMP \\
\hline
\hline
\end{tabular}
\end{center}
\end{table*}

\begin{figure*}
    \centering
    \includegraphics[width=0.45\linewidth]{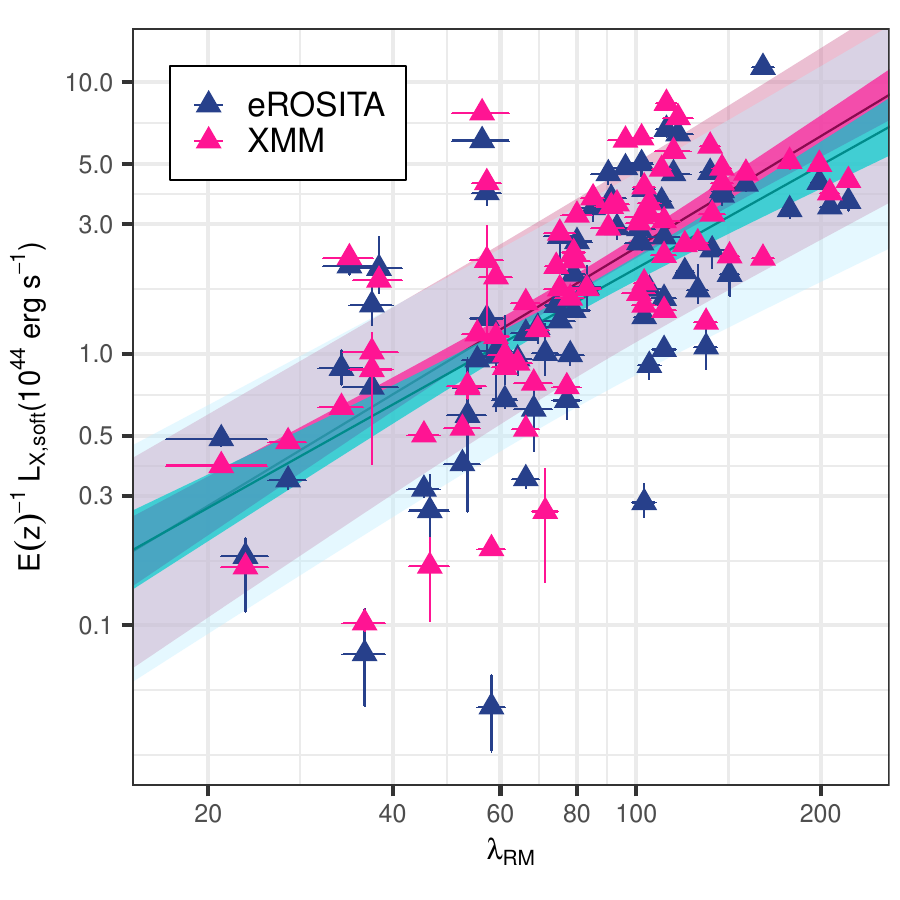}
    \includegraphics[width=0.45\linewidth]{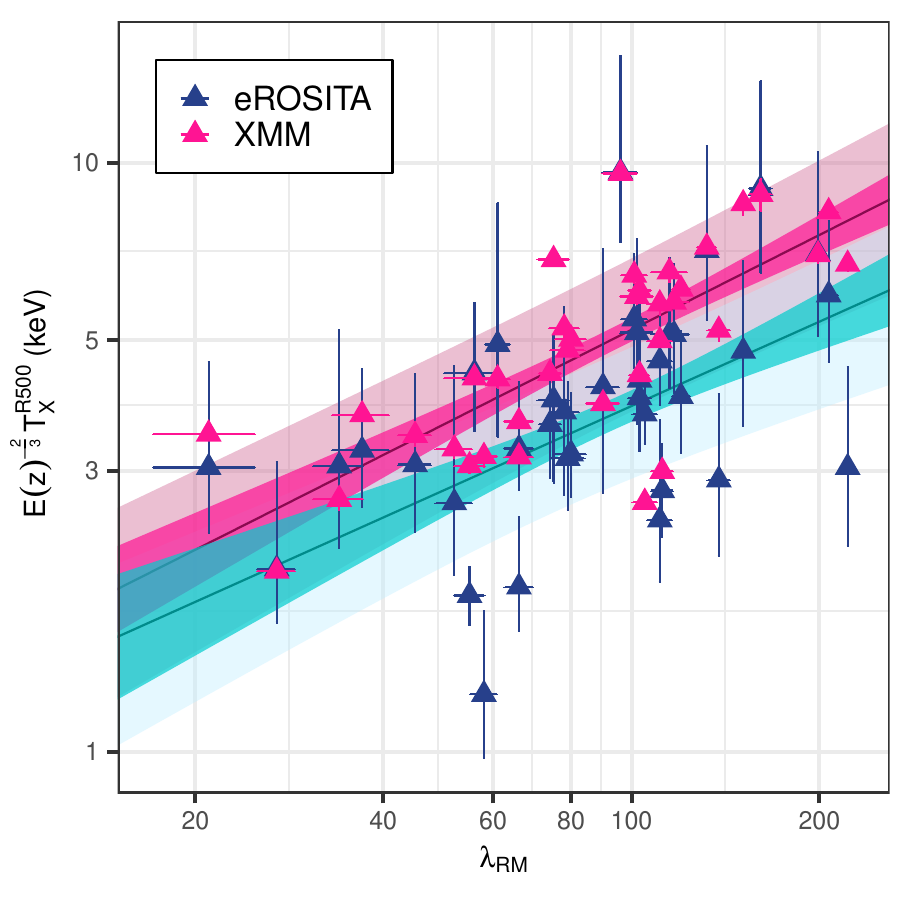}
    \caption{The $L_{\rm X, soft}$-$\lambda_{RM}$ relation (left) and the $T_{\rm X}$-$\lambda_{\rm RM}$ relation (right), each for the same cluster sample, measured with \xga{} for \erosita{} (blue) and \xmm{} (pink). The dark lines show the best-fit relations, with the surrounding cyan (\erosita{}) and pink (\xmm{}) bands indicating the 1$\sigma$ uncertainty on the fits. The light shaded regions correspond to the 1$\sigma$ intrinsic scatter.}
    \label{fig:richness_scalrel}
\end{figure*}

The left panel of Figure~\ref{fig:richness_scalrel} shows the $L_{\rm X, soft}$-$\lambda_{RM}$ scaling relation using \xmm{} (pink points) and \erosita{} (blue points). \xmm{} and \erosita{} measure consistent slopes of \xRICHLsoftSLOPE{} and \eRICHLsoftSLOPE{} respectively. \erosita{} measures a slightly lower normalisation than \xmm{} (\eRICHLsoftNORM{} compared to \xRICHLsoftNORM{}), which is expected given that \erosita{} measures $L_{\rm X, soft}$ 19$\pm$6\% less than \xmm{} (Table~\ref{tab:tel_comp}). The measured scatter is similar, with \xmm{} and \erosita{} measuring  $\sigma=$ \xRICHLsoftSCATTER{} and  \eRICHLsoftSCATTERT{} respectively.  
Similar to Section~\ref{subsec:tx_crosscal}, we repeated the scaling relation fit for $L_{\rm X, soft}$-$\lambda_{RM}$ using the subsample of \RICHTSAMP{} clusters with well constrained $T_{\rm X}$. We find no significant difference in the fit parameters between the well constrained $T_{\rm X}$ subsample and the full $L_{\rm X, soft}$ subsample. 

The right panel of Figure~\ref{fig:richness_scalrel} shows the $T_{\rm X}$-$\lambda_{\rm RM}$ relation.  
 \xmm{} and \erosita{} measure consistent slopes of \xRICHTSLOPE{} and \eRICHTSLOPE{} respectively.
\erosita{} measures a lower normalisation than \xmm{} (\eRICHTNORM{} compared to \xRICHTNORM{}), which is expected given that \erosita{} measures $T_{\rm X}$ 28$\pm$7\% less than \xmm{} (Table~\ref{tab:tel_comp}).
The measured scatter is similar, with \xmm{} and \erosita{} measuring  $\sigma=$ \xRICHTSCATTER{} and  \eRICHTSCATTER{} respectively.  

These results highlight that the choice of observatory does not have a significant impact on the slope or scatter of the derived X-ray - richness relations. To our knowledge, this study is the first to compare X-ray - optical scaling relations between \xmm{} and \erosita{}. However, \cite{Kelly2024} performed a similar comparison between \xmm{} and \chandra{}. They found that the $T_{\rm X}$-$\lambda_{\rm RM}$ relation exhibits a steeper slope when measured with \xmm{} than with \chandra{} 
(0.61$\pm$0.04 versus 0.43$\pm$0.09), along with a marginally smaller intrinsic scatter (0.23$\pm$0.02 and 0.27$\pm$0.03 respectively). Hence, it is not unexpected that our measured relations are instrument dependent. That said, our \xmm{} $T_{\rm X}$-$\lambda_{\rm RM}$ slope (\xRICHTSLOPE{}) is shallower than that of \cite{Kelly2024} and of \cite{Giles2022SDSS} (0.61$\pm$0.04 and 0.59$\pm$0.04 respectively). In principle, the scaling relations derived from a single instrument should be consistent. However, in practice the measured relations are a strong function of the selection function.
Our \xmm{} relation is derived from a sample that is additionally required to be detected by \erosita{}. Owing to the shallower depth of \erass{} relative to the \xmm{} observations, this selection preferentially includes the brightest clusters at a given $\lambda_{RM}$, particularly at low richness. These systems are expected to have systematically higher $T_{\rm X}$, and hence $L_{\rm X}$, values, which acts to flatten the measured slope \citep{Mantz2010}. Future releases of \erosita{} photon data should alleviate this bias, because longer exposure times will allow the detection of fainter systems. We note that the scatter on the $T_{\rm X}$-$\lambda_{\rm RM}$ relation measured in this study is consistent with that measured by \cite{Kelly2024}, \xRICHTSCATTER ~vs  $0.23\pm0.02$. 

In summary, this analysis has shown that the slopes and scatter of scaling relations derived from two different telescopes (\erosita{} and \xmm{}) from the same clusters are consistent. This gives us confidence that it will be possible to combine clusters detected by only one of the telescopes to improve the statistical accuracy, extend the mass range, and explore evolution in the scaling relations (assuming a well defined selection function). This is because, generally speaking, the \xmm{} archive probes to lower masses and/or higher redshifts than \erass{}, due to its longer exposure time, whereas \erass{} delivers more of the rare high mass systems, due to its larger footprint. We note that even though the normalisation of the scaling relations will be offset between the telescopes, due to the effects described in Section~\ref{sec:XMMCalib}, but this can be compensated for by scaling, as was done in \cite{Kelly2024} (where \chandra{} $T_{\rm X}$ measurements were scaled to \xmm{} for joint fits).





\section{Discussion}
\label{sec:Dis}

\label{subsec:xmmcompDis}
The cross-calibration between \xmm{} and \erosita{} $L_{\rm X, bol}$ and $L_{\rm X, soft}$ measurements (see Section~\ref{sec:XMMCalib}) shows that \xmm{} consistently measures higher $L_{\rm X}$ values than \erosita{}, with the discrepancy being larger (34$\pm$7\% offset) for $L_{\rm X, bol}$ compared to $L_{\rm X, soft}$ (19$\pm$6\% offset).  A similar trend is seen for $T_{\rm X}$, where \xmm{} measures 28$\pm$7\% higher $T_{\rm X}$ values than \erosita{}.

Two arguments are typically invoked to explain systematic measurement offsets between X-ray observatories. First the effect of modelling a multi-temperature gas with a single-temperature model. Second, calibration discrepancies in the effective area calculations. We explore these two arguments below.

\subsection{Effect of a multi-temperature ICM}
\label{sec:disc:multiICM}
This study uses a \texttt{constant$\times$tbabs$\times$apec} model to fit spectra, which assumes that the ICM is a single temperature plasma. This is an oversimplification; it is well known that many clusters have cool cores and/or temperature structures in their outskirts (e.g. \democluster{}, Figure~\ref{fig:projtempprof}). However, due to the signal to noise limitations of \erass{} observations, it is not practical to try to fit to a more complex model.
Given the differing shape of the \erosita{} and \xmm{} effective areas, each telescope will be sensitive to different components of a multi-temperature gas for a given cluster. Notably, the effective area of \erosita{} above \textasciitilde 2 keV drops significantly compared to that of \xmm{} \citep[see Figure 9 of][]{erosita}. Therefore, fits to \erosita{} spectra, under the assumption of a single temperature ICM, will be biased by the cooler components of a multi temperature ICM.  Therefore, we can expect \erosita{} to measure both lower global  temperatures and luminosities than \xmm{}. 
Although \erosita{} has slightly larger sensitivity in the 0.5–2 keV band, \xmm{} measures higher $L_{\rm X, soft}$ values because $L_{\rm X}$ is proportional to the integrated model (which is fitted over 0.3–7.9 keV), so it is influenced by \xmm{}’s high-energy response. 
The offsets will be larger for $L_{\rm X, bol}$, than for $L_{\rm X, soft}$, because the bolometric band includes the energy range where \xmm{}’s effective area exceeds \erosita's. 

Observational biases of spectroscopic $T_{\rm X}$ measurements arising from thermal structures in the ICM propagate into biases in hydrostatic mass. As shown in \cite{Rasia2014, Henson2017, Pearce2020, Barnes2021}, spectroscopic $T_{\rm X}$ measurements differ from the true thermal temperatures of simulated clusters, and when used to calculate hydrostatic mass, biases of up to 40\% are induced.

Other cross-calibration studies such as \citetalias{Migkas2024} and \cite{Lopez2025} run  $T_{\rm X}$ comparisons excluding the projected core, where the most significant multi-temperature structure along the line-of-sight is expected. 
\citetalias{Migkas2024} provide a comparison of the temperature cross-calibration in two annular regions: a core region, and an outer annulus excluding the core. In the core region, the line of sight is intersected by multiple gas temperatures, within an outer annulus, only the outskirts of the cluster is considered. \citetalias{Migkas2024} finds there is no significant difference in the cross-calibration between the core and outer annulus temperatures.  Furthermore, \citetalias{Migkas2024} simulate a range of cluster spectra consisting of cool and hot gas components, which is then fit by a single temperature model.  Performing this exercise for \erosita{}, \xmm{} and \chandra{}, \citetalias{Migkas2024} find that while \erosita{} measures consistently lower temperatures, the difference compared to \xmm{} and \chandra{} is always $\lesssim20\%$.  

In this work we do not exclude the core, since the sample does not have sufficient counts to do so, and still measure a reliable $T_{\rm X}$ value. However, the \citetalias{Migkas2024} results described above demonstrate that the cross-calibration differences found in in Section~\ref{sec:XMMCalib} cannot be fully explained by the use of single temperature fits to multi-temperature gas.

\subsection{Calibration of the effective area}
\label{sec:bandDis}

If the ICM has a single-temperature, but measured parameters are offset between different observatories, then it is possible that calibration of the respective effective areas is the cause.
It is standard practice for X-ray cross-calibration studies of clusters to repeat $T_{\rm X}$ fitting across soft and hard energy bands. Doing so can give an insight into which energy ranges are driving the calibration inconsistencies. 
We therefore repeated spectral fitting using the same sample explored in Section~\ref{sec:XMMCalib}, using the 0.4 - 5.0 keV (soft) and 1.5 - 7.0 keV (hard) bandpasses (as performed in \citetalias{Migkas2024}). 
The fit to the soft band, yielded a full set (37) of successful$^{20}$  $T_{\rm X}$ measurements, we measure an offset of 30$\pm$10\% at an \erosita{} $T_{\rm X}$ of 3 keV and a slope of 1.0$\pm$0.2, consistent with the offset and slope obtained from the full 0.3-7.9 keV band. Our results are comparable to \citetalias{Migkas2024}, who found that soft and full band $T_{\rm X}^{CE}$ measurements agree within 1$\sigma$. 

Unfortunately, the fit to the hard band yielded only 11 successful$^{20}$  $T_{\rm X}$ measurements, insufficient to constrain a reliable offset. The number of sources with measurable $T_{\rm X}$ values in this band is limited by the declining sensitivity of \erosita{} above $\sim$2 keV, leading to low signal in the hard band. As a result, the spectral fits lack sufficient constraining power to yield model parameters with uncertainties meeting our quality criterion of a fractional uncertainty ($\frac{\Delta T_{\rm X}}{T_{\rm X}}$) less than 0.8. 
This lack of constraint in the hard band means we are unable to draw conclusions regarding potential calibration uncertainties in the observatories. It may be possible to do so after future releases of eRASS photon data, these will improve the signal to noise of the \erosita{} observations of the  DES-XCS-eRO sample.

In their hard-band (1.5 - 7.0 keV) analysis, \citetalias{Migkas2024} similarly found that fewer clusters yielded reliable $T_{\rm X}^{CE}$ measurements with \erosita{}. However, they did report an offset of 55\% at an \xmm{} $T_{\rm X}$ of 3 keV (see their Figure 3). This was based on a poorly constrained hard band relation, i.e. a slope of $0.442 \pm 0.205$, compared to $0.825 \pm 0.074$ for their full-band relation, but \citetalias{Migkas2024} were able to make the conclusion that the \erosita{}–\xmm{} cross-calibration was better in the soft band and for cooler clusters than in the hard band and for hotter clusters, attributing this to the effective area calibration of \erosita{} being better matched to \xmm{} at low energies. 


\section{Conclusion}
\label{sec:conc}
In this work we present the first application of the \xga{} and \daxa{} software packages to \erosita{} data. We have validated our methods using both pointed and survey \erosita{} observations. First, we leveraged the \erosita{} CAL-PV pointed observation of  \democluster{}. We measured radial profiles of temperature with \xga{} for \democluster{} that are consistent in shape with those from \citetalias{Whelan2022}. We also showed that derived profiles, including gas density, and hydrostatic mass, show close agreement between \erosita{} and \xmm{}. Second, we used comparisons to \citetalias{Migkas2024} and \citetalias{Bulbul2024} to demonstrate that the \xga{} method provides robust measurements of the following when \erosita{} is operated in survey mode: cluster temperature within \rad{} ($T_{\rm X}$), annular temperature from 0.2 - 0.5 \rad{} ($T_{\rm X}^{CE}$), and luminosity across the 0.5 - 2.0 keV band ($L_{\rm X, soft}$).

We then applied \xga{} to a sample of 114 Dark Energy Survey redMaPPer galaxy clusters that have been detected in X-ray by both \xmm{} and \erosita{} (the DES-XCS-eRO sample). 
Of these 114, $L_{\rm X, soft}$ values were measured by \xga{} for 76 (\erosita{}) and 105 (\xmm{}) clusters. An additional cluster had an \xmm{} luminosity measured across the 0.1 - 10.0 keV band $L_{\rm X, bol}$. 
Modelling the offsets using a power-law with a fixed slope, 
we reveal that \xmm{} measures $19\pm6$\% and $34\pm7$\% higher $L_{\rm X, soft}$ and $L_{\rm X, bol}$ values than \erosita{} respectively. Allowing the slope of the power-law to vary indicates a $L_{\rm X}$-dependent offset: \xmm{} is $18\pm5$\% higher at $L_{\rm X,soft} = 2\times10^{44}\ \mathrm{erg\ s^{-1}}$, and consistent with \erosita{} at $8\times10^{44}\ \mathrm{erg\ s^{-1}}$. For $L_{\rm X, bol}$, offsets range from $50\pm10$\% at $2\times10^{44}\ \mathrm{erg\ s^{-1}}$ to $26\pm8$\% at $8\times10^{44}\ \mathrm{erg\ s^{-1}}$.

Of the 114 DES-XCS-eRO clusters, \erosita{} and \xmm{} provided $T_{\rm X}$ measurements for 47 and 103 clusters respectively. Applying a 150-count minimum for \erosita{} left \SAMPT{} clusters suitable for cross-calibration. Comparisons of the $T_{\rm X}$ measurements show that \xmm{} measures $28\pm7$\% higher values on average. Unlike M24, no evidence of a $T_{\rm X}$-dependence in the calibration was found.

Scaling relations of $L_{\rm X, soft}$-$\lambda_{RM}$ and $T_{\rm X}$-$\lambda_{\rm RM}$ were fitted for both observatories. The $L_{\rm X, soft}$-$\lambda_{RM}$ relation shows minimal difference between \erosita{} and \xmm{}, while the $T_{\rm X}$-$\lambda_{\rm RM}$ normalisation differs significantly: \eRICHTNORM{} (\erosita{}) versus \xRICHTNORM{} (\xmm{}), though slopes are consistent within $1\sigma$ (\eRICHTSLOPE{} vs \xRICHTSLOPE{}). The scatter on the  $L_{\rm X, soft}$-$\lambda_{RM}$ relation is larger than that of the $T_{\rm X}$-$\lambda_{\rm RM}$ relation (by at least a factor of 2).

We have thus demonstrated that cluster cosmology studies can leverage X-ray measurements from either observatory, without requiring individual clusters to be detected by both. Moreover, we have shown that $L_{\rm X}$ measurements  -- which require lower signal-to-noise than $T_{\rm X}$ measurements -- can be used to constrain the mass observable relations. This is particularly relevant for samples of clusters with significant overlap with \erass{} (from which $L_{\rm X}$ measurements are readily available, but $T_{\rm X}$ measurements are not). For example, samples derived from wide area optical/near-IR photometric surveys such as the DES Y3 redMaPPer catalogue, which has $\sim$1,700 \erass{} and $\simeq$325 \xmm{} corresponding detections (with the DES-XCS-eRO sample representing the overlap between these two samples). Our findings are also applicable to cluster samples detected via the Sunyaev-Zel’dovich Effect. For example the 10,040 clusters in the Atacama Cosmology Telescope’s (ACT) DR6 \citep[][]{2026OJAp....955863A}, of which 1,904 have a match in the \erass{} cluster catalogue and 712 have a match to an extended XCS source, with 251 clusters in common. 

\section*{Acknowledgements}
We thank the anonymous referee for their helpful comment regarding the choice of abundance table.

PG, KR, RW, JP, and recognise support from the UK Science and Technology Facilities Council via grants ST/T000473/1 and ST/X001040/1 (PG, KR), ST/X508822/1 (JP), ST/P006760/1 (RW). DT is supported by NASA under award number 80GSFC24M0006. MH acknowledges financial support from the National Research Foundation.

This work is based on data from eROSITA, the soft X-ray instrument aboard SRG, a joint Russian-German science mission supported by the Russian Space Agency (Roskosmos), in the interests of the Russian Academy of Sciences represented by its Space Research Institute (IKI), and the Deutsches Zentrum für Luft- und Raumfahrt (DLR). The SRG spacecraft was built by Lavochkin Association (NPOL) and its subcontractors, and is operated by NPOL with support from the Max Planck Institute for Extraterrestrial Physics (MPE). The development and construction of the eROSITA X-ray instrument was led by MPE, with contributions from the Dr. Karl Remeis Observatory Bamberg \& ECAP (FAU Erlangen-Nuernberg), the University of Hamburg Observatory, the Leibniz Institute for Astrophysics Potsdam (AIP), and the Institute for Astronomy and Astrophysics of the University of Tübingen, with the support of DLR and the Max Planck Society. The Argelander Institute for Astronomy of the University of Bonn and the Ludwig Maximilians Universität Munich also participated in the science preparation for eROSITA. 

The eROSITA data shown here were processed using the eSASS software system developed by the German eROSITA consortium.

\section*{Data Availability}
The XMM data underlying this article were accessed from the XMM science archive at \url{https://nxsa.esac.esa.int/nxsa-web/}.

The \erosita{} cluster catalogue used in this work is publicly available at \url{https://erosita.mpe.mpg.de/dr1/AllSkySurveyData_dr1/Catalogues_dr1/}, and the 
underlying \erosita{} data are accessible via the eROSITA-DE Data Release 1 
portal at \url{https://erosita.mpe.mpg.de/dr1/erodat/}.

The redMaPPer catalogue used in this paper is publicly available at \url{https://des.ncsa.illinois.edu/releases/y3a2/Y3key-cluster}.

The data underlying the results in this article are available in the accompanying GitHub repository found at \url{https://github.com/jessicapilling/eRO-XMM_Cross_Calibration}.



\bibliographystyle{mnras}
\bibliography{main} 



\appendix
\section{Masking the Cool Gas Extension in \democluster{}}
\label{app:splodge}

\begin{figure*}
  \centering
  \begin{subfigure}{0.45\textwidth}
    \centering
    \includegraphics[width=\linewidth]{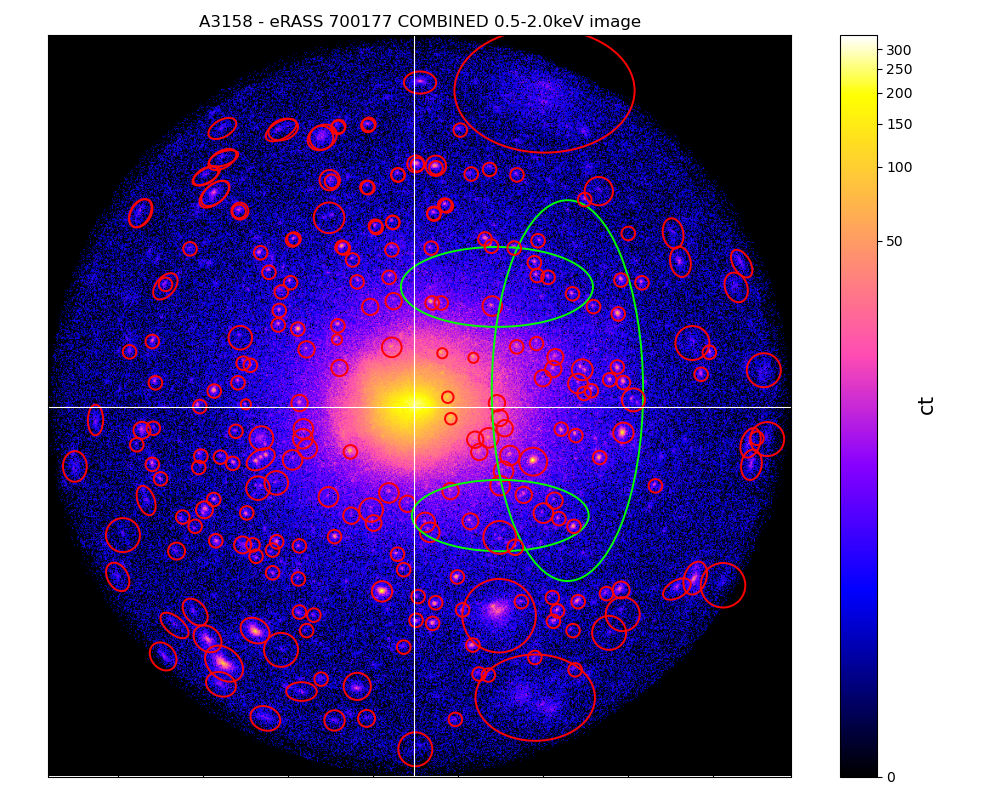}
  \end{subfigure}
  \hfill
  \begin{subfigure}{0.45\textwidth}
    \centering
    \includegraphics[width=1.\columnwidth]{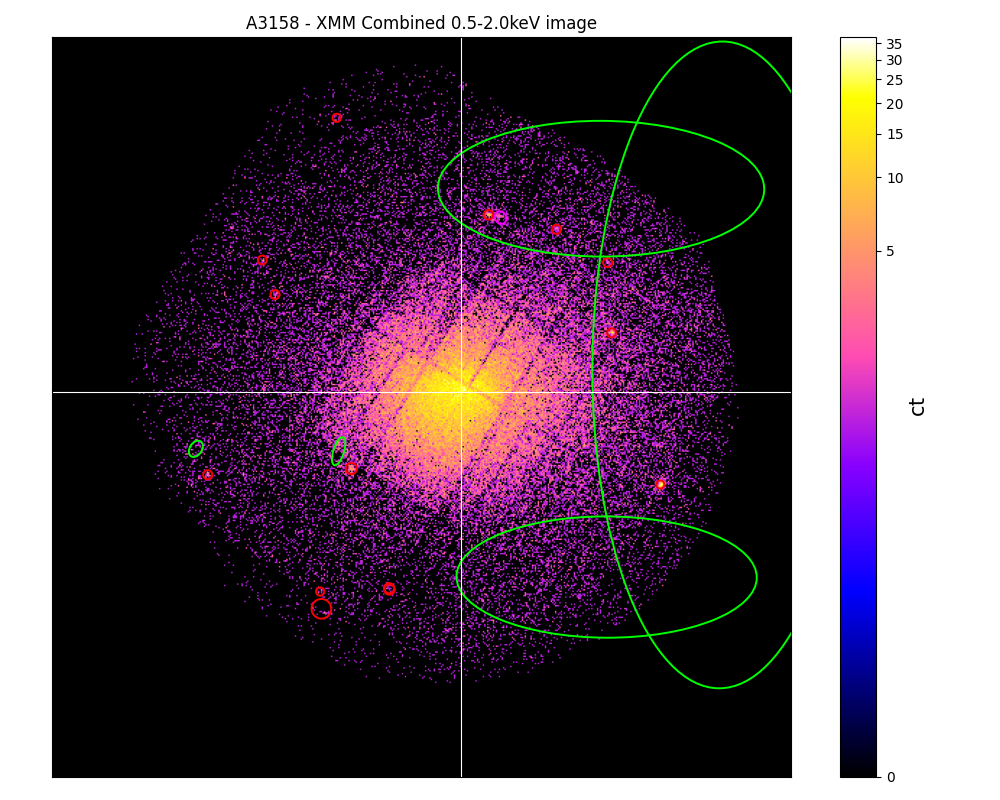}
  \end{subfigure}
 \caption{Masking used in \erosita{} and \xmm{} temperature profile generation (seen in Figure~\ref{fig:tempprof_splodge}), where the proposed cool gas extension location is approximately masked in the overlapping green ellipses.}
 \label{fig:splodge_mask}
\end{figure*}

The initial comparison to \citetalias{Whelan2022} in Figure~\ref{fig:projtempprofw22} shows consistent agreement with \erosita{} except between \textasciitilde600 - 900 kpc. This is the same region as the cool gas extension reported in \citetalias{Whelan2022}. To investigate whether this cool gas was driving the difference in the temperature profile at \textasciitilde600 - 900 kpc, this region was masked with three ellipses applied to \erosita{} and \xmm{} data, as seen in Figure~\ref{fig:splodge_mask}.

\begin{figure}
    \centering
    \includegraphics[width=1.\columnwidth]{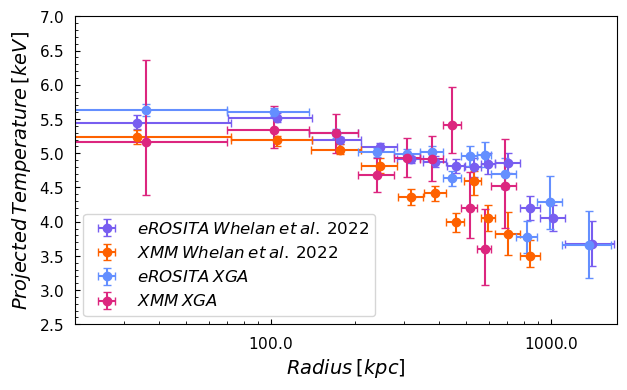}
    \caption{Projected temperature profile of \democluster{} measured with \xga{} for \xmm{} (pink) and \erosita{} (blue), compared with previous measurements by \protect\citetalias{Whelan2022} for \xmm{} (orange) and \erosita{} (purple). The equivalent to Figure~\ref{fig:projtempprof}, except masking the cool gas extension using the masks shown in Figure~\ref{fig:splodge_mask}.}
    \label{fig:tempprof_splodge}
\end{figure}

The same analysis, whereby the default \xga{} method was altered to best match the \citetalias{Whelan2022} methodology was used to measure a temperature profile for \erosita{} and \xmm{}. These results are shown in Figure~\ref{fig:tempprof_splodge}, where all \erosita{} points now appear consistent with \citetalias{Whelan2022}.

\bsp	
\label{lastpage}
\end{document}